\documentclass[article,twocolumn,pra]{revtex4-1}
\usepackage[utf8]{inputenc}
\usepackage{tikz}
\usetikzlibrary{shapes.geometric}
\usetikzlibrary{decorations.pathreplacing,calligraphy}
\usepackage{amsmath}
\usepackage{amsfonts}
\usepackage[caption=false]{subfig}

\begin{document}

\title{A quantum-inspired method for solving the Vlasov-Poisson equations}

\author{Erika Ye}
\email{eyerik.a@gmail.com}
\affiliation{Plasma Science and Fusion Center, Massachusetts Institute of Technology, 77 Massachusetts Av., Cambridge, MA 02139}

\author{Nuno F. G. Loureiro} 
\email{nflour@psfc.mit.edu}
\affiliation{Plasma Science and Fusion Center, Massachusetts Institute of Technology, 77 Massachusetts Av., Cambridge, MA 02139}

\date{\today}

\begin{abstract}
Kinetic simulations of collisionless (or weakly collisional) plasmas using the Vlasov equation are often infeasible due to high resolution requirements and the exponential scaling of computational cost with respect to dimension. Recently, it has been proposed that matrix product state (MPS) methods, a quantum-inspired but classical algorithm, can be used to solve partial differential equations with exponential speed-up, provided that the solution can be compressed and efficiently represented as an MPS within some tolerable error threshold. In this work, we explore the practicality of MPS methods for solving the Vlasov-Poisson equations in 1D1V, and find that important features of linear and nonlinear dynamics, such as damping or growth rates and saturation amplitudes, can be captured while compressing the solution significantly. Furthermore, by comparing the performance of different mappings of the distribution functions onto the MPS, we develop an intuition of the MPS representation and its behavior in the context of solving the Vlasov-Poisson equations, which will be useful for extending these methods to higher dimensional problems.
\end{abstract}

\maketitle

\section{Introduction}
Understanding the behavior of collisionless plasmas would greatly further our research of astrophysical phenomena and fusion energy systems. The Vlasov equation, a 6-D nonlinear partial differential equation (PDE), provides an {\it ab-initio} description of the dynamics of such plasmas and is deemed to be the gold standard in plasma simulation. 
It can be solved deterministically using Eulerian \cite{Valentini2007hybrid, Cerri2017kinetic, palmroth2018vlasov, vonAlfthan2014vlasiator, Juno2018discontinuous} or semi-Lagrangian \cite{Cheng1976integration, sonnendrucker1999semilagrangian, crouseilles2010conservative, liu2010conservative, liu2021conservativeFEM, Einkemmer2020semilagrangian, Kormann2019massively,  Kormann2015semilagrangian, Rahn2022parallel6D} grid-based methods. Unlike the alternative particle-in-cell (PIC)  approach~\cite{Dawson1983particle, Fonseca2002osiris, Franci2018camelia, Franci2018solar, Qin2015canonical, Xiao2021expliciit, Xiao2021symplectic, Sun2015explicit}, these methods do not suffer from stochastic noise issues; however, they are extremely computationally expensive thanks to the exponential scaling of cost with respect to dimensionality and the resolution requirements stemming from the multi-scale dynamics that characterizes the vast majority of nonlinear plasma behavior. 
These issues seriously limit our ability to simulate collisionless plasma phenomena, thus hindering progress in a wide range of fundamental and applied problems. 

In this paper, we investigate an alternative approach: the use of matrix product states (MPS) to solve the Vlasov-Poisson equation within a finite-difference scheme.
Matrix product states are a quantum-inspired computational framework traditionally employed in the simulation of quantum many-body systems, where they have been used with great success \cite{Schollwock2011dmrg, Vidal2003efficient, White1993density, White2005density, Chan2002highly}. However, it has recently been proposed that the utility of MPS methods extends beyond quantum applications, and that one can use MPS to solve PDEs with (formally) exponential reduction in computational cost \cite{Ripoll2021quantuminspired, Lubasch2018multigrid}.

An MPS is an ansatz that provides an approximate but systematically improvable low-rank representation of the data of interest. Furthermore, the MPS framework also provides a means of efficiently manipulating the data within this representation. 
Formally, one choice of MPS ansatz is equivalent to the tensor train representation, in which the data is decomposed into a series of tensors each corresponding to one of its dimensions, and then compressed by limiting the rank (the correlations) between each dimension. Tensor trains have been employed to solve PDEs in a variety of contexts ranging from fluid dynamics to molecular electronic structure 
\cite{Dolgov2012fast, Boelens2020tensorBGK, Einkemmer2019lowrankcompressible, Rakhuba2016gridbasedelectronic}, including the Vlasov-Poisson and Vlasov-Maxwell equations in up to 6-D space \cite{Dolgov2014lowrank, Kormann2015semilagrangian, Rahn2022parallel6D, Ehrlacher2017adaptive, Einkemmer2018lowrankprojectorsplitting, Einkemmer2020lowrankprojectorsplitting}. 
However, the intended MPS ansatz mirrors that of \textit{quantized} tensor trains \cite{Khoromskij2011quantics, Oseledets2009approximationLogparmas, Oseledets2010approximationQTT}, in which the data is decomposed into smaller components such that one can limit the correlations \textit{within} each dimension as well. To the best of our knowledge, quantized tensor trains have only been discussed in a limited number of contexts including solving the Fokker-Planck equation \cite{Dolgov2012fast}, the chemical master equation \cite{Kazeev2013lowrank, Kazeev2014direct}, and finite element solvers of elliptic multi-scale problems \cite{Kazeev2016qttfem, Kazeev2018quantizedfem, Kazeev2020quantizedfem}. More relevant is the recent work by Gourianov {\it et al.}, in which they demonstrate the efficiency of MPS methods for simulating Navier-Stokes turbulence in two and three dimensions~\cite{Gourianov2021quantuminspired}. Still, the physics of the Navier-Stokes equation (a fluid equation) is fundamentally different from that of the more precise Vlasov equation (a kinetic equation) discussed here. Thus, our detailed investigation of MPS methods in the context of the Vlasov equation is novel and warranted.

If the solution to a PDE can be efficiently represented as an MPS, meaning that the rank required to represent the solution within some tolerable error is roughly logarithmic with respect to the size of the data, then the computational cost of solving the PDE would also scale polylogarithmically with respect to the size of the data, formally achieving exponential speed-up over classical direct numerical simulation methods. Our work serves as an exploratory investigation into the so-called compressibility of the solutions to the Vlasov-Poisson system and the practicality of using MPS methods to solve for its dynamics. While we only consider systems with one coordinate in space and one coordinate in velocity (1D1V), we are able to draw conclusions about the efficiency of the representation and discuss considerations for scaling up to higher dimensions.

This paper is organized as follows. We first provide a brief introduction to matrix product state (MPS) algorithms, though we point the reader to Refs.~\cite{Ripoll2021quantuminspired} and~\cite{Lubasch2018multigrid} for a more thorough introduction. 
We then present our results, starting by first investigating the efficiency with which the MPS ansatz can represent the solutions to the Vlasov-Poisson equations. After verifying that the MPS ansatz is indeed an efficient representation, we investigate the practicality of solving these equations completely within the MPS framework, which involves performing compressions (i.e.,~low-rank approximations) of the state at each time step.
We conclude with an analysis of our results and a discussion of future work.

\section{Problem Statement}

The Vlasov equation describes the evolution of the distribution function of particle species $s$, $f_s(\textbf{x},\textbf{v},t)$, over the phase space defined by position ($\textbf{x}$), velocity ($\textbf{v}$), and time ($t$). 
In the presence of electric forces only, it is given by
\begin{align}
    \frac{\partial f_s}{\partial t} + \textbf{v}_s \cdot \nabla_{\textbf{r}} f_s + 
    \frac{q_s}{m_s} \textbf{E} \cdot \nabla_{\textbf{v},s} f_s = \mathcal{C}[f_s],
\end{align}
where 
$q_s$ and $m_s$ are the the species charge and mass, respectively. The operator $\nabla_\textbf{r}$ denotes the gradient in spatial coordinates, and $\nabla_{\textbf{v},s}$ denotes the gradient taken along the velocity coordinates, which can be discretized differently for the ions and electrons. 
The electric field $\textbf{E}$ is only defined on spatial coordinates; in the electrostatic case that we consider here, it is computed from Poisson's equation, 
\begin{align}
    \nabla^2 \phi & = - \frac{1}{\varepsilon_0} \sum_s q_s \int_{-\infty}^{\infty} f_s\, d\textbf{v}, \\
    \textbf{E} & = -\nabla \phi,
\end{align}
where $\phi$ is the scalar electric potential and $\varepsilon_0$ is the permittivity.
The term $\mathcal{C}[f_s]$ is the collision operator, which is assumed to be small for high temperature plasmas. In this work, if not specified, we take this term to be zero. Otherwise, we use the Dougherty collision operator \cite{dougherty1964model, lenard1958plasma}.  

We choose to solve the Vlasov-Poisson equation using finite differences on a uniform grid in real space and analyze three paradigmatic test cases: nonlinear Landau damping, the Buneman instability, and shock wave formation. Details on the set-up of these problems can be found in the Methods section at the end of the paper.

\section{Method Overview}
\subsection{MPS Representation of Classical Data}

Suppose we can represent $f$, the solution to our $K$-dimensional PDE, on a discretized grid with $N=d^L$ grid points along each dimension, resulting in a total of $d^{KL}$ data points. 
We can equivalently represent the data as a $KL$-legged tensor for which the size of each dimension is $d$, or \[ f \left(x_1,\hdots,x_K \right) \cong f \left(i_1,\hdots,i_{KL} \right), \] where the set indices $\{i_j\}$ can take on integer values from 0 to $d-1$ and index the position of the element of interest along the $j^\text{th}$ leg. (To avoid confusion with the dimensionality of the PDE ($K$), we refer to the dimensionality of a tensor as its number of legs; the origin is from tensor network diagrams, 
in which an $n$-dimensional tensor is represented as a shape with $n$ legs sticking out of it.)

\begin{figure}
    \centering
    \subfloat[]{\includegraphics[width=0.32\linewidth, trim={0cm 0 0 0}, clip]{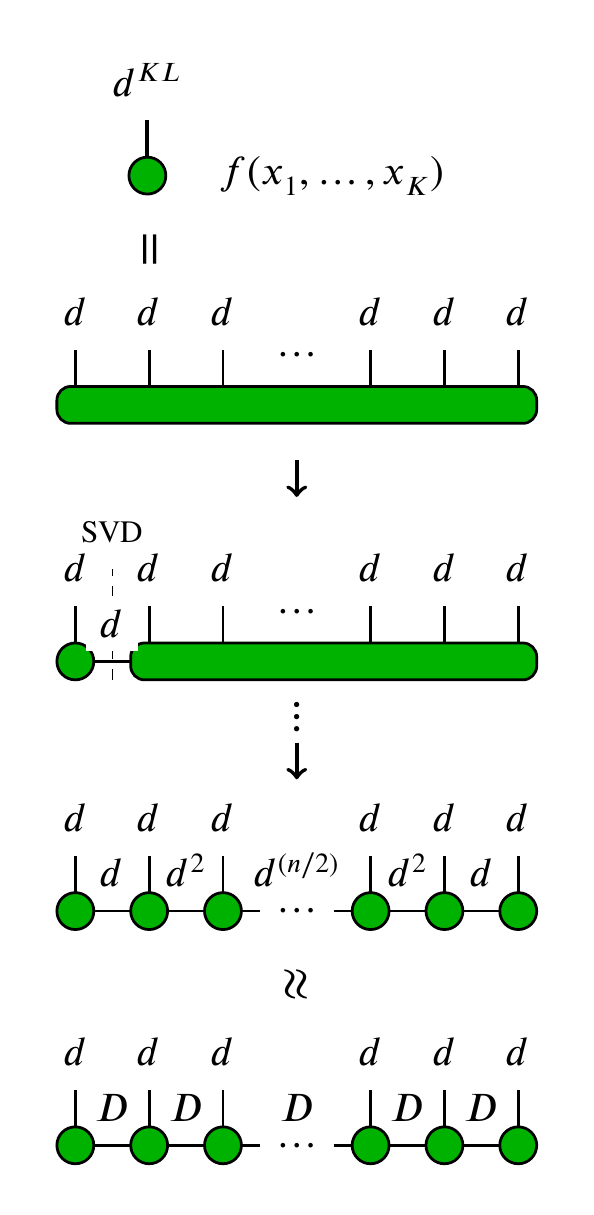}}
    \subfloat[]{\includegraphics[width=0.68\linewidth, trim={0cm 0cm 0.25cm 0}, clip]{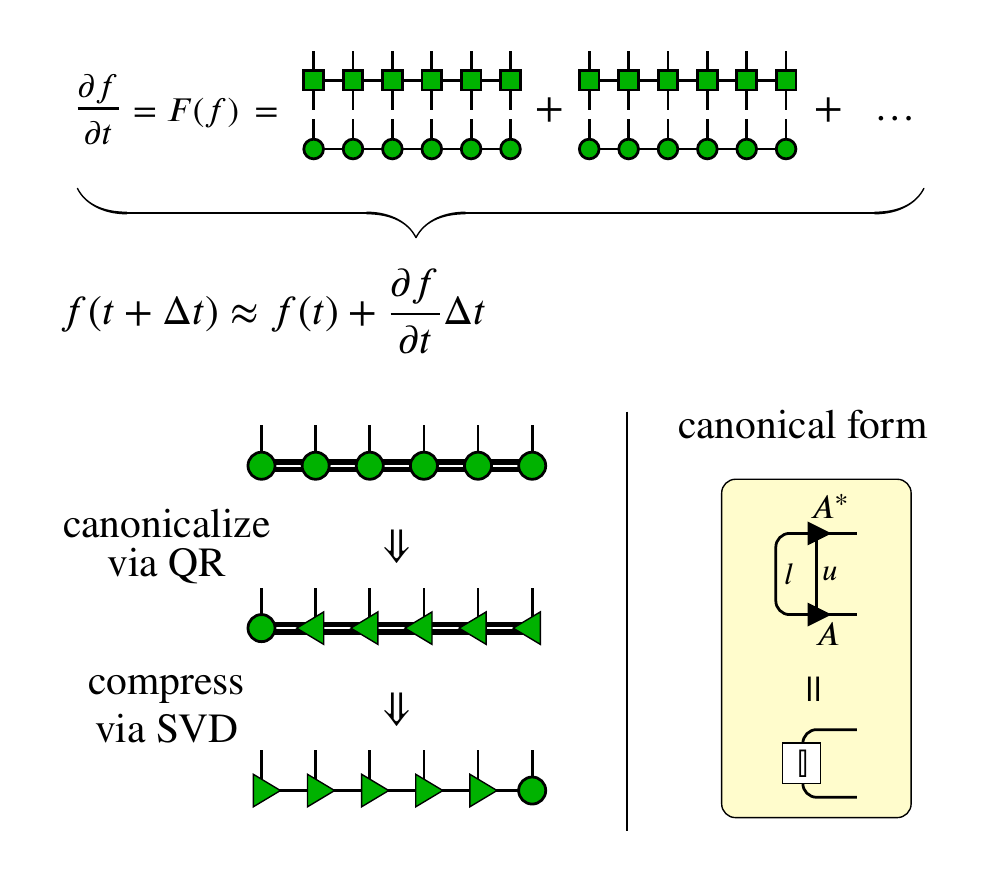}}
    \caption{Tensor network diagrams depicting (a) the conversion of a vector into an MPS, and (b) solving the PDE using MPS. We first compute the time derivative, assumed to be a function of our state $f(t)$, within the MPS framework. This involves applying the desired operations (represented as matrix product operators (MPOs), depicted as 1-D chains of square tensors) to the state and summing all of the terms together. We then compute the state at the next time step. At this point, the bond dimension of $f(t+\Delta t)$ is larger than its original value due to the various operations applied to the MPS, so the MPS needs to be recompressed. This is done by first canonicalizing the MPS using QR decompositions, and then performing the compression via SVD and retaining only the $D$ largest singular values at each bond (see SI for details). In these diagrams, $n$-dimensional tensors are represented by shapes with $n$ legs. Legs that are connected to each other represent tensor contractions along those dimensions. Tensors in canonical form are depicted using triangles and obey the property denoted in the yellow box, or that $\sum_{u,l} A_{l,r}(u) A^*_{l,r'}(u) = \delta_{r,r'}$. }
    \label{fig:svd_decomp}
\end{figure}

We then decompose this tensor into an MPS by performing singular value decompositions (SVDs) in a iterative fashion, yielding
\begin{align}
    f \left(i_1,\hdots,i_{KL} \right) = \hspace{-1cm} & \nonumber \\
    \sum_{\alpha_1=1}^{r_1} & ... \sum_{\alpha_{KL-1}=1}^{r_{KL-1}} \hspace{-0.1cm} 
    M^{(1)}_{\alpha_1}(i_1) M^{(2)}_{\alpha_1,\alpha_2}(i_2) \hdots M^{(KL)}_{\alpha_{KL-1}}(i_{KL}) 
\end{align}
where $M^{(j)}$ are 3-legged tensors (2-legged for $j=1$ and $j=KL$) and $r_{j}$ is the rank associated with the SVD decomposition between tensors $j$ and $j+1$. This decomposition is depicted in Fig.~\ref{fig:svd_decomp}(a).

For an exact, full-rank representation, as one travels towards the middle of the chain, the rank increases exponentially, with $r_j = d^{\min(j,KL-j)}$. 
However, 
one can obtain a low-rank approximation of the original state by only retaining components corresponding to the $D$ largest singular values in the decomposition at each bond. 
%
%
%
The error arising from compression at bond $j$, defined as the normalized Frobenius norm of the difference between the original state $f$ and the compressed state $f_D$, is 
\begin{equation}
    \epsilon_j = 
    \frac{\left| \left|f-f_D \right| \right|_F }{ \left|\left| f \right| \right|_F }
    = \left( \sum_{i=D}^{r_j} \left( \tilde{\sigma}^{(j)}_i \right)^2 \right)^{1/2} \, ,
\end{equation}
where $\{ \tilde{\sigma}^{(j)} \}$ are the singular values at bond $j$, ordered from largest to smallest, normalized such that $\sum_i ( \tilde{\sigma}_i^{(j)} )^2 = 1$. Note that the singular values must be obtained when the MPS is in the proper canonical form (see SI for details). 
The parameter $D$, often referred to as the bond dimension, thus determines both the accuracy of the representation and the computational cost of the MPS algorithm. The compressibility of the data, such that it can be represented as an MPS of some small bond dimension (to be defined later) with tolerable error, is the crux of MPS algorithms.

\subsection{Time Evolution Using MPS}

One advantage of the matrix product framework is that one can perform linear operations on an MPS while remaining in matrix product form, eliminating the need of converting it back to its original vector form. 
Fig.~\ref{fig:svd_decomp}(b) depicts the procedure of performing time evolution with MPS.
Analogous to the representation of vectors as MPS, operators that act on the state $f$, such as the gradient, can also be written in matrix product form (ie. matrix product operators, or MPOs). Elemental multiplication can be reframed as multiplication of the state with a diagonal operator, and taking the dot product involves taking the sum of those products. Details on these operations are in the SI.
Solving the initial value problem is then no different from traditional matrix-vector multiplication methods, except all the calculations are done with the vectors and matrices in matrix product form.

However, as one performs operations on the state $f$, the bond dimension will grow, eventually becoming unnecessarily large and unmanageable. Thus, one needs to compress the MPS back to the desired bond dimension. The compression of the data is done by first putting the MPS into canonical form via iterative QR decompositions and then performing the low-rank approximation using iterative SVD decompositions (see the SI and Ref.~\cite{Schollwock2011dmrg} for details). Unfortunately, the compression step does introduce computational overhead, which we discuss in the next section.

\subsection{Cost Analysis}
\label{section:cost}

Assume that our state $f$ is represented as an MPS with bond dimension $D$, and we are operating on the state with an MPO of bond dimension $D_w$. The cost of applying the MPO to the MPS is simply the cost of tensor contraction, which scales like $\mathcal{O}(KL D_w^2 D^2 d^2)$. The resulting MPS now has a bond dimension of $D D_w$, and we wish to compress it to bond dimension $D$. 
The cost of putting the MPS in canonical form via QR decomposition scales like $O(d D^3 D_w^3 KL)$, and the cost of the actual SVD compression scales like $\mathcal{O}(d^2 D^3 D_w KL)$. As such, the cost of the MPS algorithm is dominated by the canonicalization and compression of the MPS at each time step. If $D_w$ is constant and independent of the number of grid points in the discretization, which is the case for the linear finite difference operations on a uniform grid (see SI and Ref.~\cite{Ripoll2021quantuminspired} for details), then one can say the algorithm formally scales like $\mathcal{O}(D^3 K\log_d{N})$. In comparison, traditional sparse matrix-vector multiplication scales like $\mathcal{O}(N^K)$. Thus, if $D$ scales logarithmically with vector size, the MPS framework provides exponential speed-up.

One complication arises with the nonlinear term in the Vlasov-Poisson equation. Because the force felt by the states (i.e.~the electric field) is determined from the distribution functions themselves, the bond dimension of the MPS representation of the electric field will depend on the problem of interest and can also depend on grid size. If the bond dimension of the electric field is $D_F$, then the cost of canonicalizing the nonlinear term scales like $\mathcal{O}(D^3 D_F^3)$. 
In the worst case, $D_F$ will depend linearly with $D$. However, for the Vlasov equation, we expect $D_F$ to be closer to $\sqrt{D}$ because the electric field only has coordinates in position space and not the full phase space. In this case, the increase in computational cost mirrors that of traditional methods, which exhibit a similarly increased scaling of $\mathcal{O}(N^{3K/2})$. One potential advantage of the MPS framework is that we may be able to approximate the force term and represent it with smaller bond dimension without significant loss in the accuracy in the simulated dynamics. We elaborate on this in the Results section.

In this work, we set $d=2$. However, if we had chosen $d=N$, $L=1$, we would arrive at the more common tensor train formalism, in which the data is decomposed in between but not within each dimension.
The dominant cost of tensor train methods is also the canonicalization and compression step, which would scale like $\mathcal{O}\left( N D^3 D_w^3 K \right)$. Choosing a smaller $d$ and compressing the data within each dimension, as done here, will lead to exponential speed-up with respect to the number of grid points along each dimension, $N$. The proposed MPS method will, however, have a larger prefactor scaling the computational cost because of the bond dimensions of the operators in MPO form. In the tensor train format, operators are often separable along each dimension with $D_w = 1$. (The primary exception lies in computing the nonlinear term). Fortunately, for finite difference methods on a uniform grid, the MPO bond dimensions typically are still quite small, around 2 to 5, and independent of $N$.

While MPS methods formally show exponential reduction in computational cost, there is some overhead because of the $D^3$ scaling in the MPS compression step. 
By rough comparison of the scalings of the computational costs, we can loosely define a target bond dimension $D_\text{target}~\sim N^{K/3}$, below which MPS methods might be competitive against traditional methods. 
For $N=2^9$ and $K=2$, this would correspond to $D_\text{target}=64$, or an 8-fold reduction from the maximum possible bond dimension. For the two-dimensional (1D1V) system that we consider here, we do not expect MPS methods to significantly outperform direct numerical simulation methods. Rather, this work serves as a proof-of-principle investigation and a first step to considering higher dimensional systems in the future. 

As a brief side note, when solving the Vlasov equation, it is not uncommon to save the distribution functions with the intention of restarting the simulation from that point in time. The amount of memory required for storing an MPS of bond dimension $D$ is $\mathcal{O}(K\log_d(N)D^2d)$, which is exponentially less than the $\mathcal{O}(N^K)$ cost for storing the state in its vector representation. Since solutions to the Vlasov equation can be up to 6-D, saving them in MPS form can significantly reduce storage costs, provided that they are sufficiently compressible.

\section{Results}

The results will be presented as follows:
we start by computing the dynamics of nonlinear Landau damping without any compression (i.e.,~without low-rank approximations), and then gauge the compressibility of the electron and ion distributions within the MPS representation. We also compare results for different MPS constructions. After verifying that the distribution functions can be efficiently represented as an MPS, we then investigate solving the PDE 
by performing time evolution with compression at each time step. We present results for nonlinear Landau damping, the Buneman instability, and shock-wave formation in 1D1V. In all calculations, we use MPS with $d=2$.

\begin{figure*}
    \centering
    \includegraphics[width=\linewidth,trim={0cm 0cm 0cm 0}, clip]
    {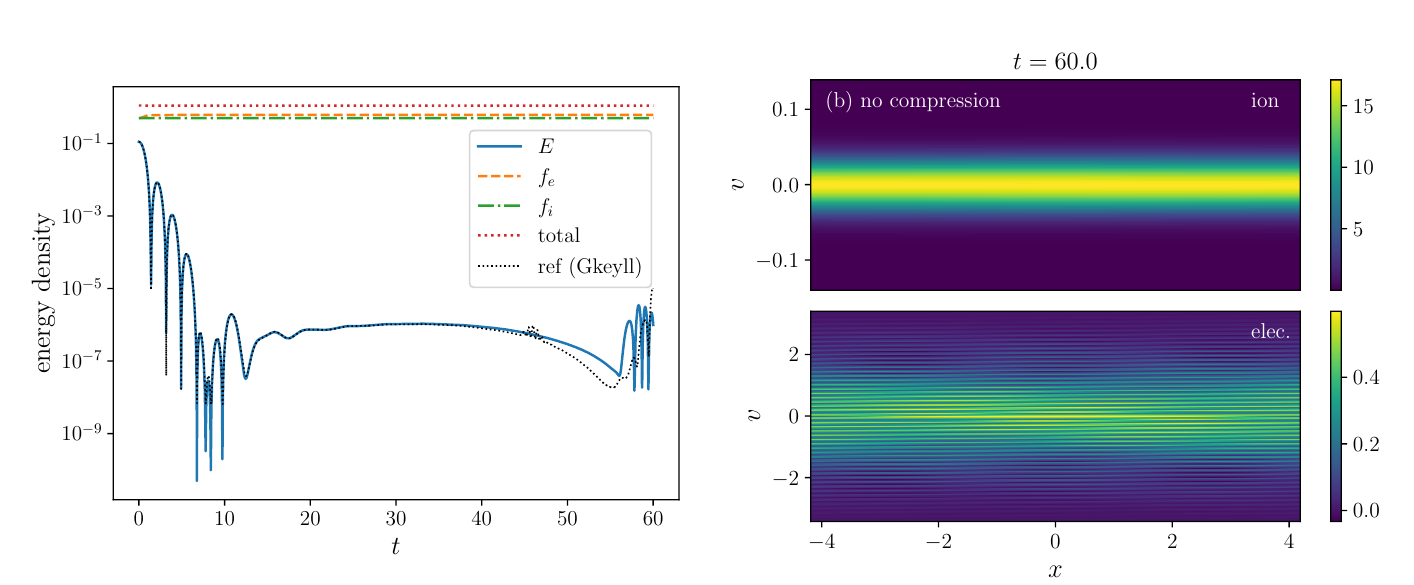}
    \caption{Landau damping without compression of the distribution functions. (a) Plot of energy density of electric field, electron distribution, and ion distribution over time for perturbation wavevector $k=0.75$. Results from Gkeyll \cite{Juno2018discontinuous} are shown in dotted black as reference. (b) The corresponding ion and electron distribution functions at $t=60$. 
    }
    \label{fig:landau_nocomp}
\end{figure*}

\subsection{State Compressibility}

Numerically exact simulations of the Vlasov-Poisson equation are performed within the MPS framework by enforcing the truncation error $\epsilon_j$ at each bond to be less than some small threshold value 
(we use a threshold of $10^{-10}$). These calculations are referred to as simulations with no compression. 
We first consider the case of nonlinear Landau damping.
As shown in Fig.~\ref{fig:landau_nocomp}, our code yields the anticipated dynamics, in good agreement with results obtained from the code Gkeyll \cite{Juno2018discontinuous} (small discrepancies are to be expected because Gkeyll is a discontinuous Galerkin finite-element code whereas we employ a (less accurate) finite-difference scheme and use a larger time step).

To quantify the compressibility of the distribution functions in the MPS representation, one can measure the Von Neumann entanglement entropy (EE) at each internal bond, defined as
\begin{align}
    S^{(j)}_\text{MPS} = - \sum_{i=1}^{r_j} ({\tilde{\sigma}_i^{(j)})^2} \log_2{ (\tilde{\sigma}_i^{(j)})^2},
\end{align}
where $\tilde{\sigma}^{(j)}$ are the normalized singular values associated with bond $j$. 
The maximally entangled case is when all $r_j$ singular values are equally weighted at $\sqrt{1/r_j}$, yielding an entanglement entropy of $\log_2({r_j})$. Generally speaking, the larger the entanglement entropy, the less compressible the state and the larger $D$ needs to be in order to accurately represent the state.

The interpretation of the entanglement entropy depends on the mapping chosen during the initial reshaping of the vector of data to the $KL$-dimensional tensor. For 1-D systems, the most straightforward choice is to have the $n^\text{th}$ data point be indexed in the MPS by the $d$-nary representation of $n$. In this mapping, the MPS has a multigrid representation, with the left-most tensor corresponding to the coarsest grid and subsequent tensors corresponding to increasingly fine grid resolutions \cite{Lubasch2018multigrid, Ripoll2021quantuminspired}.
The entanglement entropy thus measures the degree of correlations between grids of different resolutions, which in turn suggests that systems with scale separation can be efficiently represented by MPS. 

\begin{figure*}
    \centering
    \subfloat{\includegraphics[width=\linewidth, trim={-1.2cm 0.5cm -1.5cm 0, clip}]{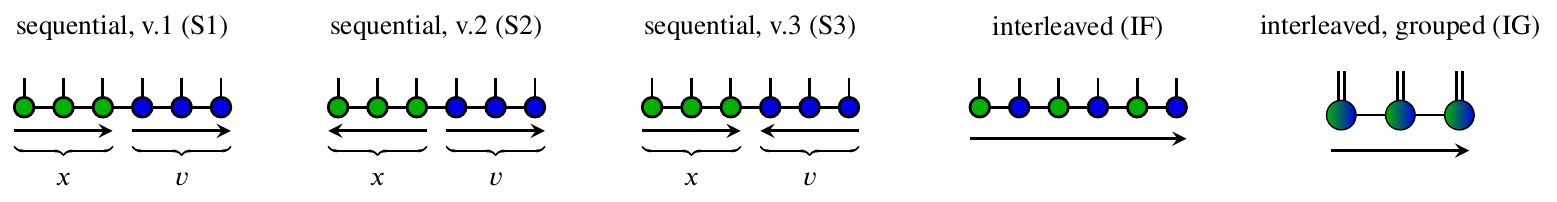}} 
    \\
    \subfloat{\includegraphics[width=\linewidth, trim={0.1cm 0cm 0.2cm 0cm},clip]
              {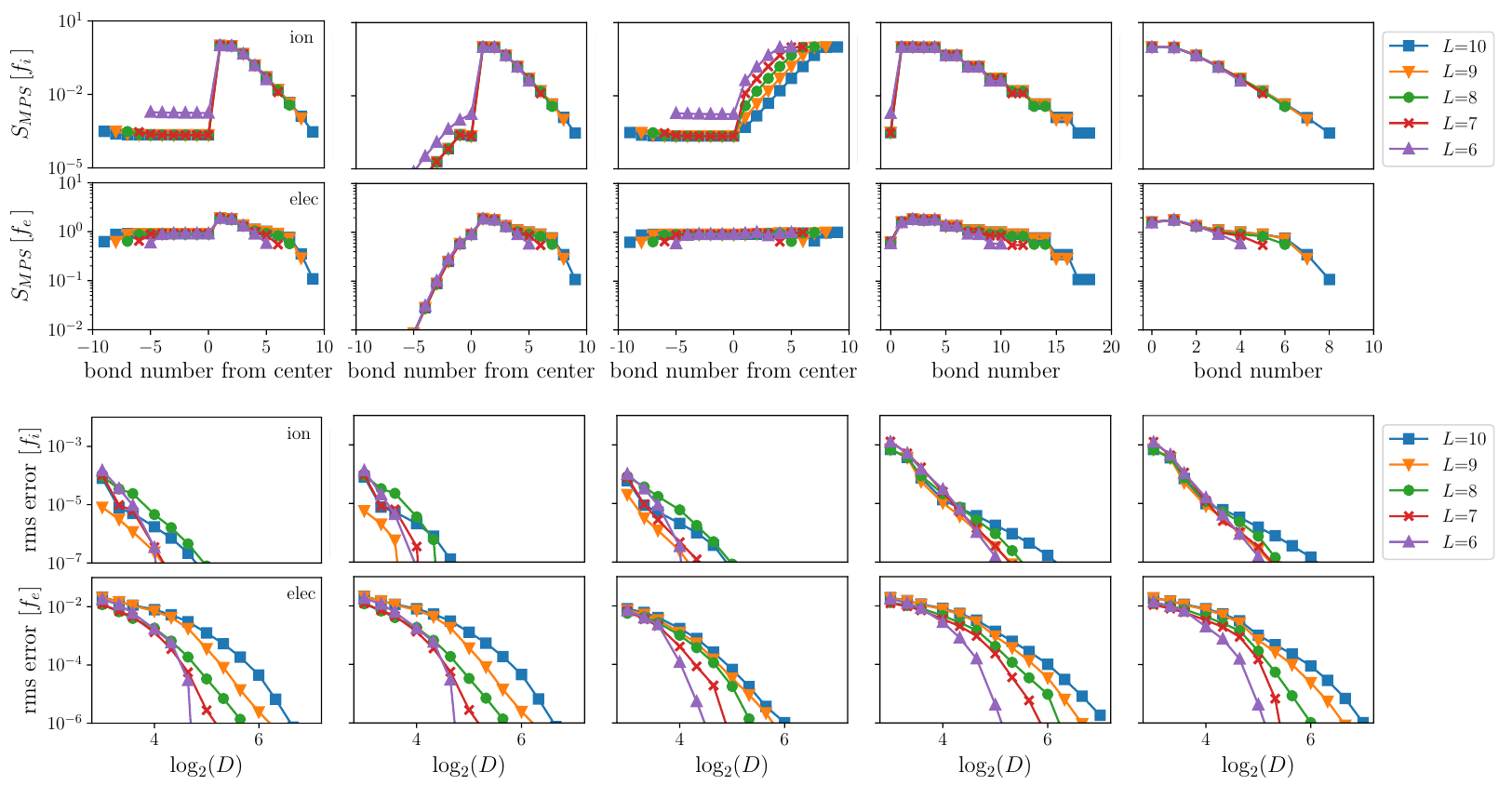}}
    \caption{Entanglement entropy for solution to 1D1V Landau damping at $t=60.0$. (top row) Different ways of ordering our 2D state in an MPS. Arrows correspond to ordering of tensors from coarse to fine grid resolution. (middle row) Entanglement entropy of the ion and electron distributions for Landau damping with initial conditions $A=0.5$ and $k=0.75$ evolved to time $t=60$ using $2^L$ grid points per dimension. 
    For the sequential orderings, the EE at the center bond corresponds to the EE if using a tensor train format. (bottom row) Root-mean-square (rms) error of the same distribution functions when compressed to bond dimension $D$ with respect to the original uncompressed representation for each grid resolution.}
    \label{fig:1_EE_k075}
\end{figure*}

For higher dimensional data, a priori, there exist multiple equally reasonable mappings one can consider. For example, our 2-D data can be mapped to the MPS such that the tensors indexing a given dimension are appended sequentially. Then, for each dimension, the tensors can then be ordered from coarse to fine grid resolution or the reverse. We consider three distinct variants in which the tensors of each dimension both go from coarse to fine (S1), the first dimension is ordered from fine to coarse while the second is ordered from coarse to fine (S2), as well as the opposite (S3). Alternatively, the data can be mapped such that tensors corresponding to similar grid resolutions but different dimensions are adjacent to each other (IF). This interleaved ordering is used by Gourianov \textit{et al.} ~\cite{Gourianov2021quantuminspired}, but they also contract tensors across different dimensions together so that now $d=2^K$ (IG). These different orderings are shown in the top row of Fig.~\ref{fig:1_EE_k075}.

The remainder of Fig.~\ref{fig:1_EE_k075}
shows the entanglement entropy of the MPS representations of the ion and electron distributions for the nonlinear Landau damping test case at the normalized time of $t=60$, as well as the root-mean-square (rms) error of the distribution function compressed to bond dimension $D$ with respect to the uncompressed result. We compare solutions obtained from solving the Vlasov-Poisson equations on grids with different resolutions ($2^L$ grid points per dimension for a fixed domain with $L$ ranging from 6 to 10), and also compare results for different MPS orderings. 

The shape of the EE curves describes the amount of correlations between different grid resolutions at each bond. The shapes change for the different MPS orderings because the different tensor orderings affect the amount of information that must be carried by each bond. However, we find that the magnitudes and spread of the EEs along each bond remain comparable for the different orderings, with the exception of the electron distribution for S3 ordering, whose maximum EE is roughly half of that of the others.

Let us first consider MPS in the S1 ordering. As shown in Fig.~\ref{fig:landau_nocomp}(b), the ion distribution appears to approximately be uniform in position space and Maxwellian in velocity space. The EE at the center bond connecting the spatial dimension and velocity dimension is on the order of $10^{-3}$ to $10^{-4}$, meaning that the distribution function is close to separable. The EE of the bonds connecting tensors corresponding to the spatial dimension is of the same order of magnitude. The small magnitude of the EE is to be expected given the simplicity of a near-uniform distribution, since there would only be negligible correlations between features of different length scales along $x$. The plateau in the EE for spatial grid tensors suggests that the EE is dominated by the correlation of the coarsest grid in $x$ to the velocity grid tensors; we verify this using the result for S2 ordering, which shows that the EEs of bonds connecting increasingly fine spatial grids to the rest of the MPS drop off by about a factor of 2. 
Because the MPS must capture the Maxwellian distribution in velocity space, the EE for bonds corresponding to the velocity dimension are nonzero. We expect the EE at coarser grid resolutions to be larger, such that it captures the general shape of the distribution function, while the EE at finer grid resolutions are smaller, since they are only responsible for adding smoothness to the function. The visible drop in the EE for increasingly fine velocity grid tensors again suggests that the dominant correlations between the $x$ and $v$ dimensions are at the coarse grid.

The electron distribution function is more interesting. The EE at the center bond is about a value of 1, meaning that the spatial and velocity dimensions are no longer separable. However, we still see a plateau in EE at finer grid resolutions in $x$, suggesting that the EE in the bonds is dominated by the correlation of velocity space with the coarsest grids in $x$. The jump in EE at the first bond from the center suggests that the additional correlations between $x$ and $v$ are mostly not with the coarse velocity grid. 
The smaller EE for the S3 ordering in this particular test case indicates that there are instead strong correlations between the coarse grid of one dimension and the fine grid of the other. Indeed, consistent with our expectations, the electron distribution (shown in Fig.~\ref{fig:landau_nocomp}(b)) exhibits slightly skewed striations along $x$. 

The same kind of analysis on the EE curves can be done for the interleaved orderings. 
For the IF ordering, we observe a step-like structure in the EE, in which the changes in the EE occur predominantly at every even bond. This means that the EE is dominated by correlations within the second dimension, which was determined above. The EE for the IG ordering has a similar shape to the IF result, meaning that there is little entanglement between the $x$ and $v$ dimensions at the paired grid scales.

We now compare the EE for simulation results obtained using different grid resolutions (different $L$).
If the distribution function is sufficiently resolved, as is the case for the ion distribution, the EE for each bond does not change when increasing the grid resolution. The bond connecting MPS to the additional tensor corresponding to the finest grid resolution exhibits an EE that appears to scale with the grid spacing. (This is because the values at these grid points can be approximated using a Taylor expansion, yielding singular values proportional to 1 and $\Delta x$ at that bond). If the distribution function is not sufficiently resolved, as is the case for the electron distribution, the EE at each bond can change. In the worst case, doubling the resolution along both dimensions would increase the EE by 1 at all bonds. However, the increase in EE observed here is much less than that, suggesting that only a small amount of information is added. For this particular test case, the additional information is encoded very efficiently for the S3 ordering, as we do not see a visible increase in EE for any of the bonds.

The entanglement entropies measured are all relatively small: the EE at the center bond has a theoretical maximum value of $L$, but the maximum observed value is 2. As such, we expect the distribution functions to be very compressible. We measure the error arising from the compression, which we define as the rms error of the compressed distribution function with respect to its uncompressed value for the specified grid resolution. Note that this is not a measure of the error of the distribution function with respect to the solution's true value.

When compressing the MPS to $D=8$, the rms errors are only on the order of $10^{-3}$ to $10^{-4}$ for the ion distribution and $10^{-2}$ for the electron distribution for all grid resolutions. Compared to the ions, the rms error for the electrons decays relatively slowly with increased bond dimension at small $D$; increasing the accuracy in the electrons from $10^{-2}$ to $10^{-3}$ requires increasing the bond dimension by a factor of 4. A slower drop in rms error suggests that the distribution function is dominated by a few modes but contains many weaker components that will need to be included in order to achieve the desired degree of accuracy. It also appears for this particular test case that the sequential ordering performs marginally better than the interleaved ordering, exhibiting smaller rms error for a specified $D$.

Comparing the bond dimension required to achieve the desired degree of accuracy provides insight on the scaling of $D$ with respect to the number of grid points along each dimension, $N=2^L$. 
For the electron distribution, if one is satisfied with relatively large compression errors on the order $10^{-3}$, the bond dimension required generally converges with increased grid resolution. However, with the exception of the S3 ordering, the bond dimension required to achieve a smaller compression error appears to increase with grid size; for a compression error on the order of $10^{-6}$, the $D$ required appears to scale like $d^{\alpha L}$ where $\alpha$ is some small constant. Because no collisions are included in these calculations, some of this additional information may be due to noise introduced at the grid level by the finite difference time-stepping scheme. 
However, as mentioned above, increasing the grid resolution does incorporate additional components weighted by coefficients that scale with grid spacing. As a result, since the compression error is measured with respect to the uncompressed solution of the specified grid resolution, the lower resolution solutions could appear more compressible since they contain less information and do not fully capture the features that can only be resolved on a finer grid. 
The convergence of rms error with respect to $L$ for the S3 ordering is in part due to the particularly low EE at all bonds for this specific test case, but also suggests that the S3 ordering is able to more efficiently represent the extra information within the higher resolution state (along with any numerical noise), such that the bond dimension required for a desired rms error converges to some finite value as one increases grid resolution.

In summary, consistent with the fact that the true distribution function only contains a finite amount of information, the EE only increases slightly as it quickly converges to that of the maximum grid resolution. This suggest that the amount of information in the distribution function can be described by the entanglement entropy at each bond of its MPS representation. As a result, the MPS ansatz should provide an efficient representation of the distribution functions, since the cost of manipulating the data is correlated to the amount of information in the state itself and not on the number of grid points in the discretization. Additionally, by comparing how the EEs vary for different MPS orderings, one can provide insight on the dominant correlations within the distribution function. 

In this nonlinear Landau damping test case, the species distribution functions can be represented with rms error on the order of $10^{-2}$ or less with small bond dimension of about $D=8$. While the rms error drops relatively slowly at small values of $D$, the curve steepens at larger bond dimension. 
The different MPS orderings appear to behave relatively similarly, though the S3 ordering yields a particularly compressible MPS representation for. The S3 ordering also exhibits the best convergence in compression error with respect to bond dimension for increasing grid resolution, such that the bond dimension required for the desired compression error does not grow linearly with the number of grid points used in the discretization.
However, note that while knowing how the bond dimension scales with respect to grid resolution is important for cost arguments, in practice, the resolution is often set by the physics.

\subsection{Compressed Time Evolution}

We now investigate the performance of MPS methods in solving the Vlasov equation while compressing the distribution functions and electric field to some prescribed bond dimension $D$ at each time step.
For the sake of simplicity, in the results presented here we only compress the MPS representing the distribution function to the specified bond dimension at each actual time step, as opposed to each intermediate time step in the standard fourth-order Runge-Kutta procedure (RK4). This is the most expensive option but also the most accurate. We refer the reader to the SI for further discussion on algorithmic variations one can consider.

In order to obtain dynamics accurate within $\mathcal{O}(\Delta t)$ of the uncompressed results at all time steps, the tolerable truncation error at each time step must be less than $\Delta t^2$ due to the accumulation of errors in the time evolution scheme. However, if one is less interested in the exact distribution function at a given point in time but more interested in the general behavior of the system, as is often the case, a larger truncation error can be tolerated. 
Unfortunately, compression can introduce numerical noise, potentially in the form of sharp features 
that would cause numerical instabilities in finite difference schemes. However, we find that when using the S3 and interleaved orderings we often can capture important features of the dynamics with remarkably robust performance, even when compressing the state by more than a factor of 8. 

\begin{figure*}
    \centering
    \includegraphics[width=\linewidth,trim={0cm 0cm 0cm 0cm},clip]{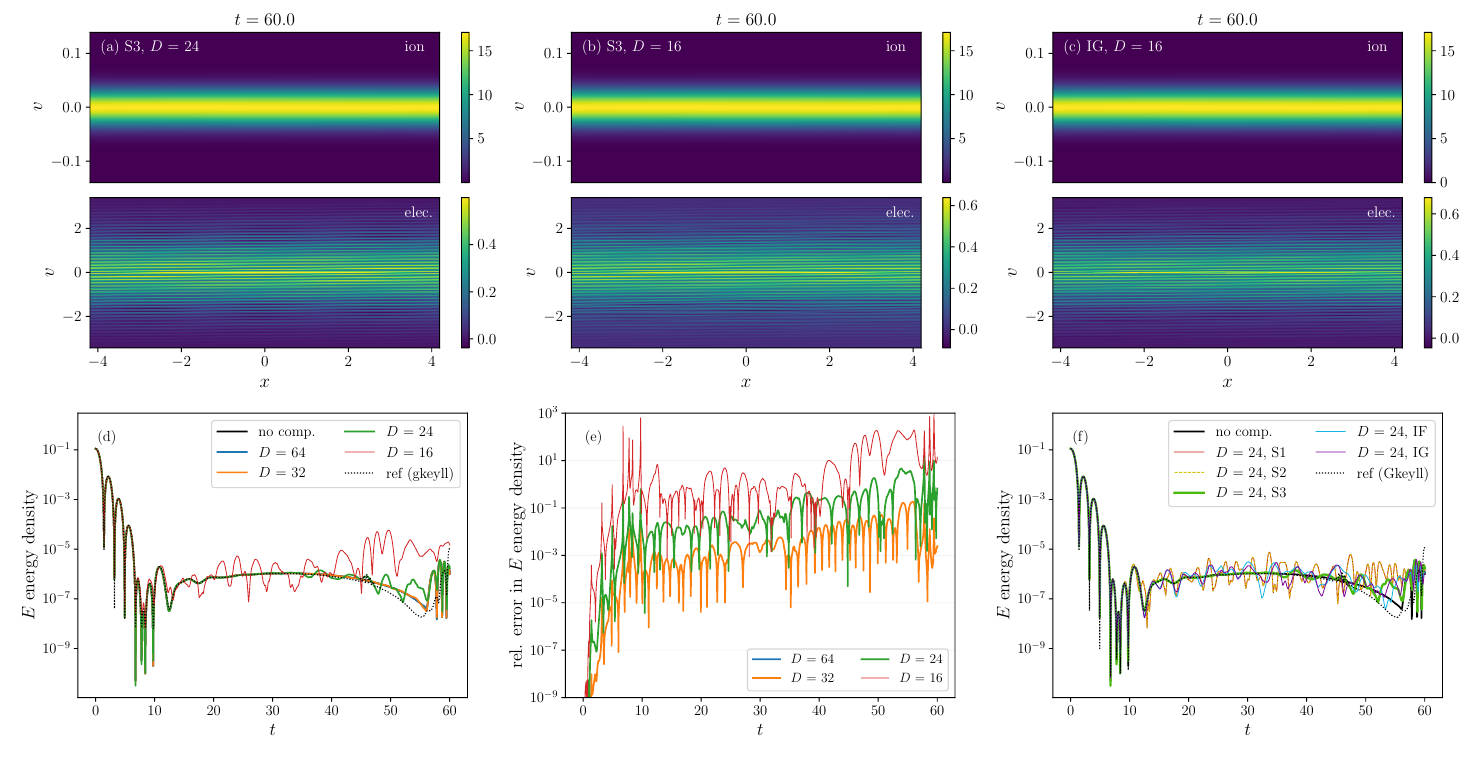}
    \caption{Compressed time evolution for nonlinear Landau damping with $k=0.75$ on a $2^6 \times 2^8$ grid, with grid spacing $\Delta x \approx 0.13$, $\Delta v_e \approx 0.047$, $\Delta v_i \approx 0.0011$. (a-c) Ion and electron distributions (over one period in $x$) at $t=60$ obtained from compressed time evolution with $D=24$ and S3 ordering, with $D=16$ and S3 ordering, and with $D=16$ and IG ordering, respectively. 
    (d) Electric field energy density for different levels of compression for MPS with S3 ordering. (e) Relative error in the electric field energy density of the compressed time evolution results with respect to the uncompressed case for different levels of compression. S3 ordering is used. Results for $D=24$ are within 10\% of the uncompressed result for times less than $t=40$. Results for $D=64$ are exact with respect to the uncompressed result, meaning that each truncation error at each bond $\epsilon_j$ is less than $10^{-10}$, and thus no blue line is shown in this plot. (f) Error in the electric field energy density of the compressed time evolution for different MPS orderings at bond dimension $D=24$.  Results for S1 and S2 ordering appear to be identical and are closely overlapping. 
    }
    \label{fig:landau_TE_k075}
\end{figure*}

\subsubsection{Nonlinear Landau damping}

We first revisit the same nonlinear Landau damping problem analyzed in previous sections. Results of solving the Vlasov-Poisson equations with compressed time evolution are shown in Fig.~\ref{fig:landau_TE_k075}.
The simulations are performed on a $2^6 \times 2^8$ grid, with grid spacings $\Delta x \approx 0.13$, $\Delta v_e \approx 0.047$, and $\Delta v_i \approx 0.0011$. The MPS representation can have a maximum bond dimension of $2^7$. 

As discussed in the Cost Analysis section, MPS methods would start to be competitive with traditional matrix-vector multiplication methods when the bond dimension is roughly $D_\text{target} \sim (N_x N_v)^{1/3}$, where $N_x$ and $N_v$ are the number of grid points used along the $x$ and $v$ dimensions. For this problem, $D_\text{target}=24$, and is about a 5-fold reduction from the maximum bond dimension of $2^7$. We find that this level of compression is very manageable, as we are still able to compute the electric field energy density within 10\% of the result obtain from uncompressed time evolution for times less than 40. For longer times, though the error grows, the energy density of the field remains close to zero---the large relative error is in part due to the small amplitude of the field. While less accurate, higher levels of compression (smaller $D$) still yield results with the correct damping and saturation behavior. Furthermore, even for the aggressively compressed $D=16$ case, the distribution functions at $t=60$ are visually remarkably similar to those of the uncompressed result (Fig.~\ref{fig:landau_nocomp}(b)), capturing the same horizontal striations with only some small differences. 

Out of the different MPS orderings considered, the S3 ordering yields the most accurate results when $D=24$. This is consistent with our previous analysis (Fig.~\ref{fig:1_EE_k075}) which showed that the S3 ordering yields the lowest maximum entanglement entropies and the smallest rms compression error. The performance of the other MPS orderings are similar to each other. Again, while these results exhibit larger error in the saturation regime, they all still capture the main features of dynamics.

\begin{figure*}
    \centering
    \includegraphics[width=\linewidth, trim={0cm 0cm 0cm 0}, clip]{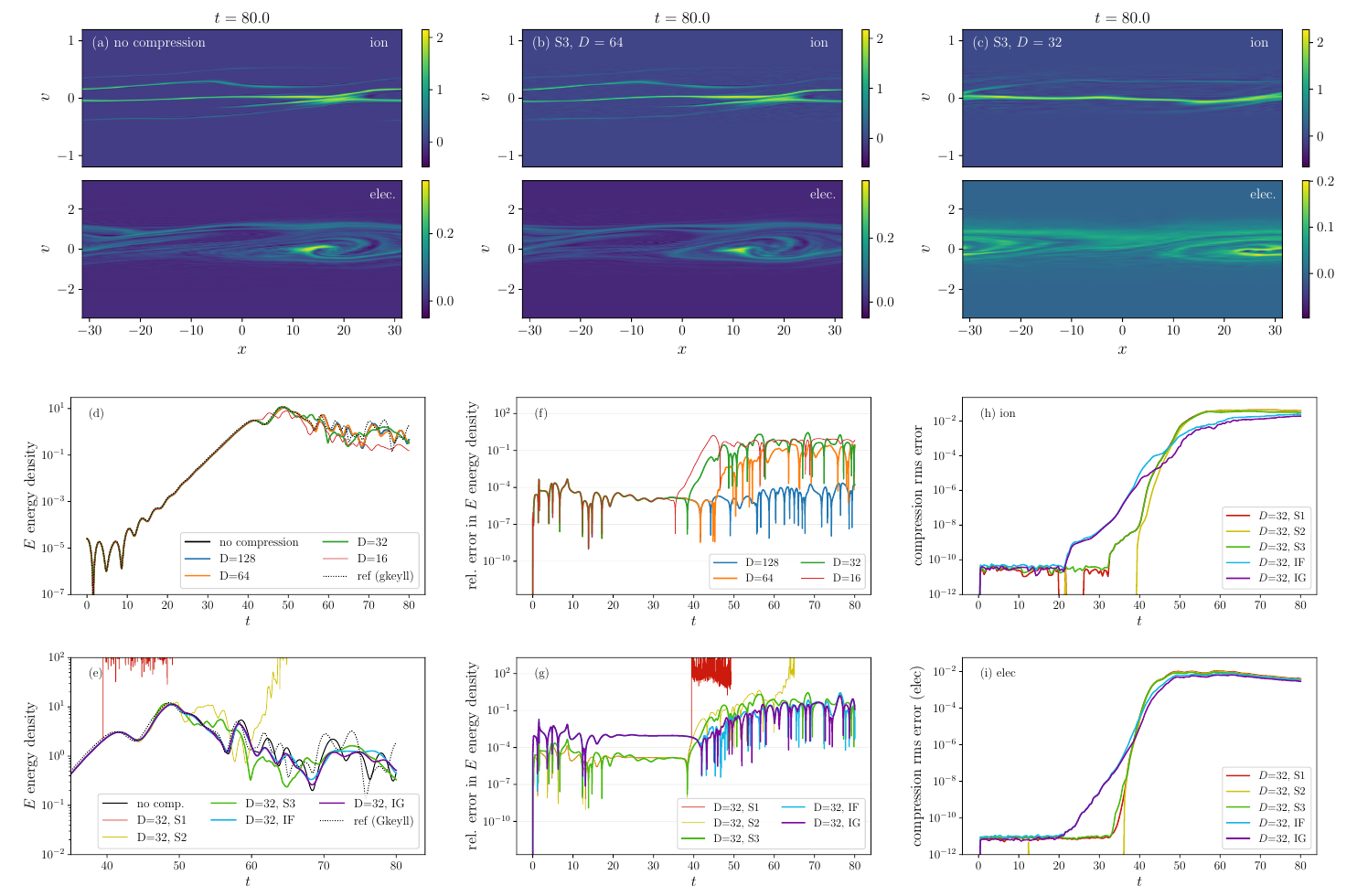}
    \caption{Buneman instability. (a-c) Distribution functions of the ions and electrons at normalized time $t=80$ obtained without compression, a bond dimension of $D=64$, and a bond dimension of $D=32$ using the S3 ordering. 
    (d) Electric field energy density for different levels of compression. The curve in dotted black is obtained using Gkeyll for an external reference.
    (e) Electric field energy density in the nonlinear regime for different MPS orderings
    (f,g) Relative error in the electric field for different compression levels and different MPS orderings. Note that the interleaved orderings have larger error in the linear growth regime but perform better in the nonlinear regime.
    (h,i) Rms error of the ion and electron distribution function after compression of the exact (uncompressed) results to $D=32$ at each time step, evaluated for different MPS orderings. 
    }
    \label{fig:3_bun}
\end{figure*}

\subsubsection{Buneman instability}
We now consider the 1D1V Buneman instability \cite{buneman1959dissipation}, for which the background configuration differs from the nonlinear Landau damping case only in that the electron distribution is now centered at $v_{0,e} = \omega_{p,e}/k$ in velocity space, where $\omega_{p,e}$ is the electron plasma frequency. For numerical reasons, we also include collisions 
with a collision frequency of about $0.007\omega_{p,s}$ for each species (the linear growth rate is about $0.2\omega_{p,e}$). The results for initial perturbation strength $A=10^{-3}$ and wavevector $k=0.10$ on a $512^2$ grid are shown in Fig.~\ref{fig:3_bun}. We find that we can compress the data to a bond dimension of 64, a factor of 8 reduction from the maximum value, and still obtain the anticipated dynamics of the system, with the energy in the electric field accurate to about 10\%. Visually, the distribution function looks similar to the uncompressed version, except for some noise in the ion distribution at small velocity magnitudes.

The complexity of the distribution functions is closely correlated to the degree of nonlinearity in the dynamics. In the linear regime at times less than about 40, the system is extremely compressible, as evidenced by small compression errors on the order of $10^{-10}$ in both the ion and electron distributions and good convergence with respect to bond dimension. However, in the nonlinear regime, the rms compression errors are at about $10^{-2}$ for a bond dimension of $D=32$. 
Consistent with our observations in the case of Landau damping, the MPS with sequential ordering perform better in the linear regime, in which the distribution function remains closer to a separable state; however, the interleaved orderings perform slightly better for the nonlinear regime in which multi-scale structures often dominate.

While the error in the electric field energy density at a given point in time for the $D=32$ result is large, we are still able to capture the general shape of the energy density over time. In Fig.~\ref{fig:3_bun}(d), we find that the energy density in the electric field is of the correct order of magnitude and roughly follows the shape of the expected results. Furthermore, as shown in Fig.~\ref{fig:3_bun}(c), we are still able to roughly see the same swirling features in the electron distribution. Interestingly, it appears as if collisions at a higher collision rate had been used since many of the features in the distribution function and the electric field energy density have been smoothed out.

\subsubsection{Approximation of the nonlinear term}

\begin{figure}
    \centering
    \includegraphics[width=\linewidth, trim={0 0 0 0}, clip]{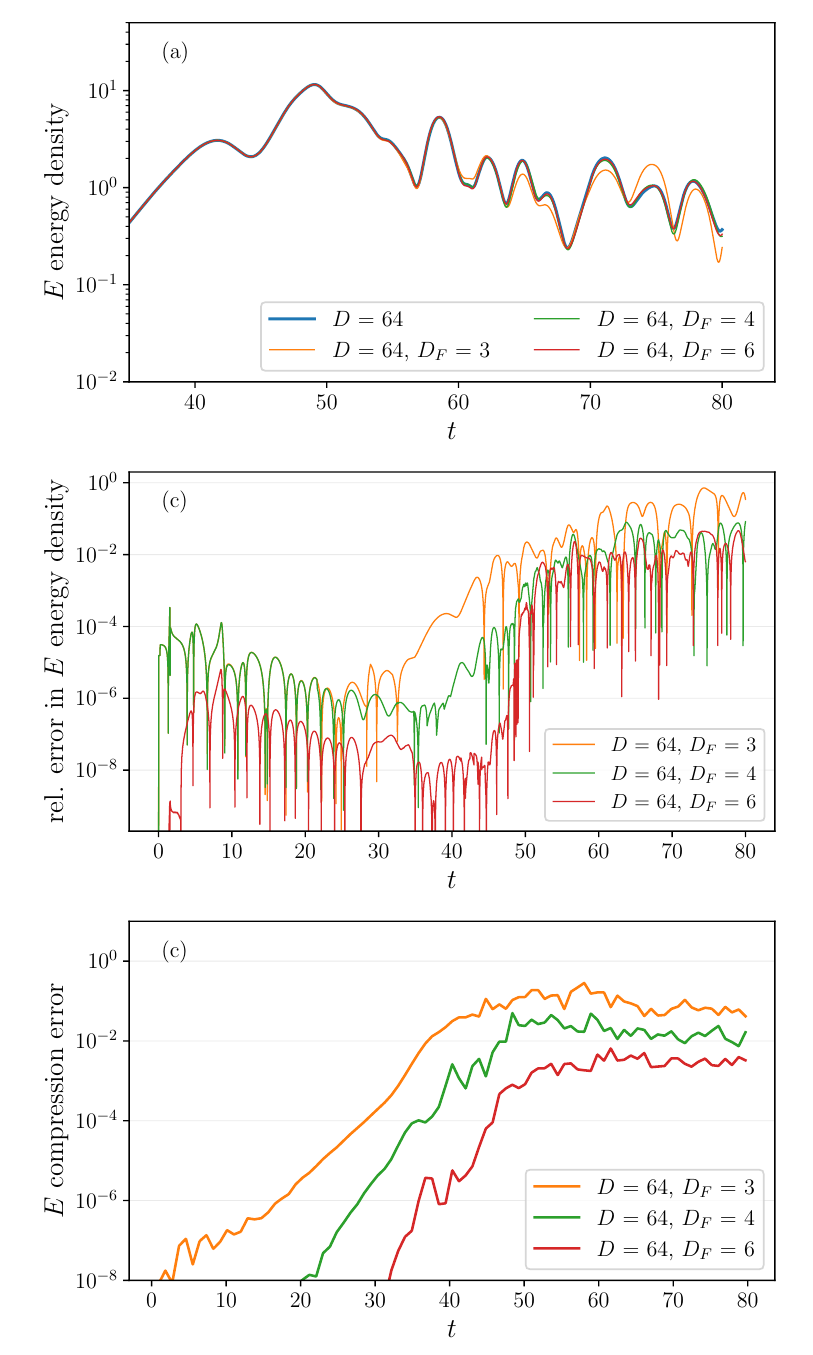}
    \caption{Approximate calculation of the nonlinear term for the Buneman instability. Ion and electron distribution functions are compressed to bond dimension $D=64$. (a) Traces of the electric field energy in time with different levels of compression of the electric field ($D_F$). 
    (b) Relative error in electric field energy density with respect to the $D=64$ time evolution result but with no compression of the electric field.
    (c) Error in the electric field after compression to the specified $D_F$ for each time step.
    }
    \label{fig:4_ED}
\end{figure}

As mentioned earlier in the Cost Analysis section, the cost of computing and compressing the nonlinear term using the MPS framework scales like $\mathcal{O}(D_F^3 D^3)$, where $D_F$ is the bond dimension of the MPS representation of the electric field. 
Since the electric field is obtained from the ion and electron distribution functions, the bond dimension of its MPS representation ($D_F$) will vary depending on the problem of interest. As such, computing the nonlinear term can potentially be significantly more expensive than the linear terms. 
However, as shown in Fig.~\ref{fig:4_ED}, when performing time evolution with the electron and ion distribution functions compressed to $D=64$, we can compress the electric field to just $D_F=4$ while still remaining within 10\% of the result obtained without compression of the electric field. The compression error of the electric field is on the order of $10^{-2}$. This shows that one is able to compress the electric field MPS aggressively without introducing significant error in the dynamics, thereby reducing the cost associated with computing the nonlinear term.
This will need to be investigated more carefully for higher dimensional problems, but it might be another source of speed up in MPS calculations. 

\subsubsection{Collisionless shocks}
\begin{figure*} 
    \centering
    \includegraphics[width=\linewidth,trim={0 0 0cm 0cm}, clip]{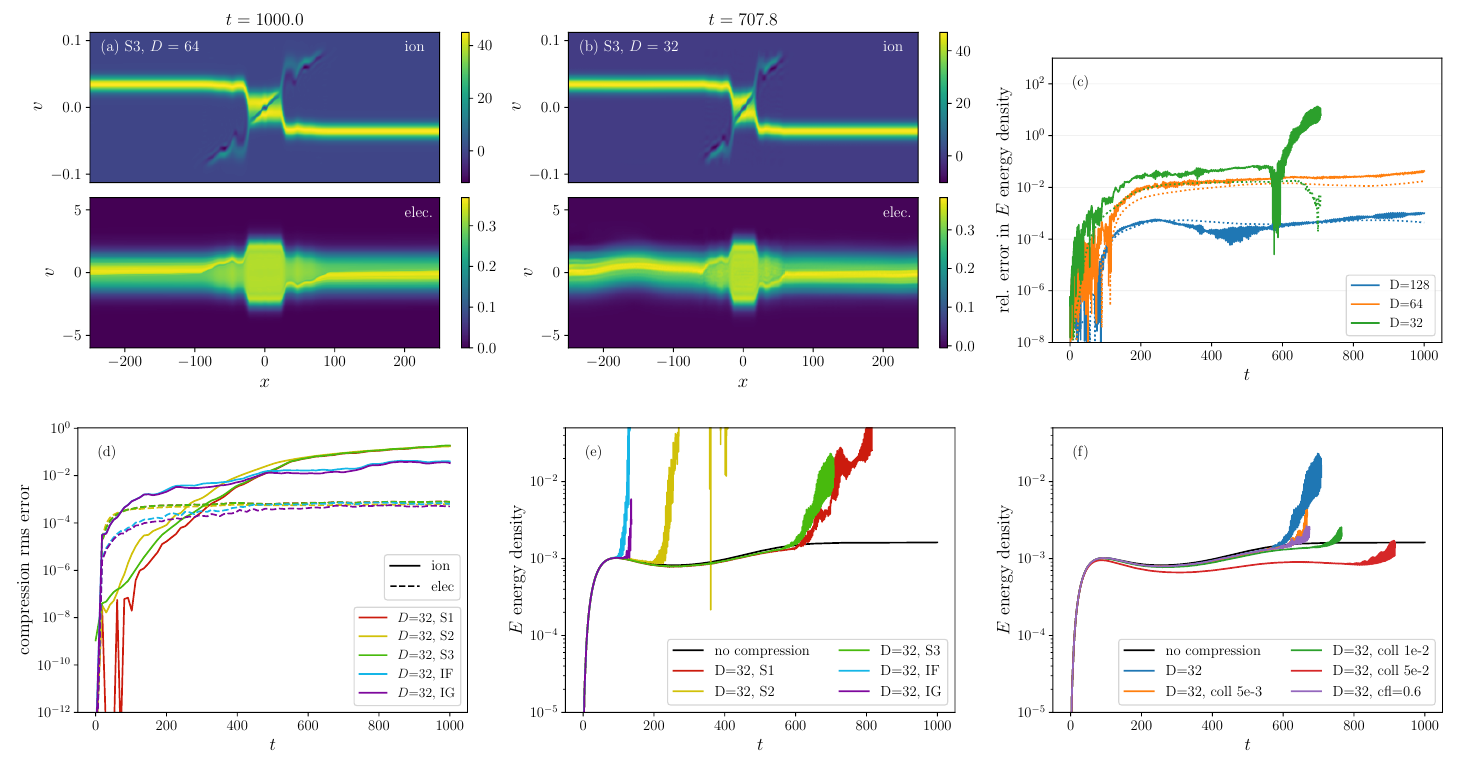}
    \caption{ Compressed time evolution for shock-wave formation on a $512^2$ grid. (a) Distribution of ions and electrons at a time of 1000 obtained using S3 ordering and bond dimension $D=64$. 
    (b) Distributions at a time of 707 obtained using S3 ordering and bond dimension $D=32$.
    (c) Relative error in electric field energy density and total energy density obtained using the specified bond dimension and S3 ordering. 
    (d) Rms error generated by compressing the uncompressed ion and electron distributions to a bond dimension of $D=32$ at each time step, evaluated for different MPS orderings. 
    (e) Electric field energy density computed using compressed time evolution with $D=32$ for different MPS orderings. (f) Electric field energy density computed using compressed time evolution, with the inclusion of various noise-mitigation methods such as adding collisions and using a smaller time step. MPS with $D=32$ and S3 ordering are used.}
    \label{fig:shocks}
\end{figure*}

Lastly, we consider the problem of shock-wave formation in plasmas. These simulations are performed on a $512^2$ grid, and the results are shown in Fig.~\ref{fig:shocks}. We are able to capture the dynamics with a bond dimension of $D=64$ within 10\% accuracy of the energy density. However, compared to the previous test cases, these calculations are much more sensitive to further compression. For example, even though the error when compressing the distribution functions to $D$=32 is less than $10^{-3}$ at early times, simulations using S2, IF, and IG orderings fail drastically during that time. MPS with S1 and S3 ordering perform better, though they also fail at about $t=600$ once the error in the ion distribution becomes on the order of $10^{-1}$.

This unstable behavior is likely due to the accumulation of compression errors over time as opposed to a numerical instability. 
When the simulation fails, the electron distribution function develops an unphysical wiggle at around $x=-150$ (see Fig.~\ref{fig:shocks}(b)). Reducing the time step does not improve performance, and while adding collisions can delay the onset of the failure, the collision rates required for stability at long times are too large as they would affect the dynamics of the system. This test case is likely more sensitive to compression errors than the previous test cases because the compression generates a finite charge density in regions where it should be zero, and this erroneous charge density may grow over time. In other words, while compression errors of the ion and electron distribution functions may be of a tolerable amount, the relative error of the charge density at each point in space is much larger.
Therefore, unsurprisingly, the MPS algorithm as presented is less advantageous for problems where high accuracy in the distribution functions is required. However, we still are able to obtain stable and reasonably accurate results with a bond dimension of $D=64$, which is just an eighth of the maximum possible bond dimension.

\section{Discussion}

The above results suggest that MPS methods can efficiently represent solutions to the Vlasov-Poisson equation in 1D1V, and that one can also use a compressed finite difference time evolution scheme to solve for the dynamics of the plasma with reduced cost while still capturing its important features. 
In most cases, we find that one can generally compress the MPS representation of the ion and electron distribution functions to a bond dimension of $D \sim d^{(L_x + L_v)/3}$ and still determine the energy of the electric field within 10\% of that obtained without any compression. We estimate this amount of compression to be enough for MPS to begin to be competitive with sparse matrix-vector multiplication methods. 
Meeting this benchmark now is encouraging, since we expect MPS methods to become even more competitive for higher dimensional systems. 
We also find that even smaller bond dimensions can be used; though less accurate, main features of the dynamics, such as growth rates and saturation energies, as well as features in the distribution function itself, can be captured. 

The compressibility of the distribution function is perhaps unsurprising, given the inefficiency inherent to finite difference methods; one needs a grid fine enough to accurately compute the gradients in the distribution function, and thus not all data points are providing significant information regarding its shape.
However, the fact remains that the MPS framework is able to 
provide a more efficient representation of the data, leading to (formally) exponentially reduced computational costs. 
Analyzing the compressibility and entanglement entropy of distribution functions when mapped to different MPS orderings (as done in Fig.~\ref{fig:1_EE_k075}) can also help provide physical insight on the dynamics, enabling us to understand how information between different grid scales and different dimensions propagates over time.

While the choice of MPS ordering did not appear to significantly affect the compressibility of a given distribution function, their performance during compressed time evolution varied significantly. For example, in the Buneman instability test case, the MPS with S1 and S2 ordering are susceptible to numerical instabilities generated by compression, whereas the interleaved and S3 orderings appear to be much more robust, even with aggressive compression. This suggests that locality of tensors corresponding to fine grids is important for the robustness of MPS methods in a finite difference scheme, because the compression tends towards removing weak features at the fine grid scales. However, robustness to noise also suggests the tendency to remove weaker fine scale features, as observed in Fig.~\ref{fig:3_bun}(c).

The errors of compressed time evolution with the S3, IF, and IG orderings were generally within an order of magnitude of each other (collisionless shocks being an exception). Nonetheless, we found that the S3 ordering performed better in the linear or weakly nonlinear regimes, in which the distribution functions largely remain separable across the two dimensions. In contrast, the interleaved orderings performed better in regimes where nonlinear effects dominate. This is within expectations, since nonlinear dynamics are known to be multi-scale and the interleaved MPS groups tensors of similar grid scale together. The grouped version (IG) appeared to yield marginally lower error than the factored version (IF). However, the computational cost of this ansatz scales like $d^K$ instead of $Kd$. While an unimportant difference for the 1D1V case, this trade-off between performance and cost would need to be investigated at higher dimensions (i.e.,~$K=6$).
Compared to the sequentially ordered MPS, the interleaved ordering may feel less natural, particularly when performing operations like derivatives and integrals along certain dimensions and when considering grids with different resolutions along each dimension. However, carefully optimized implementation aside, these are not particularly strong reasons to avoid using an interwoven ordering. Though they are less accurate than the sequentially-ordered MPS in the linear regime, such calculations often only require modest bond dimension so the compression error is still relatively small. Thus, this is a relatively small trade-off compared to increased compressibility in the nonlinear regime. 

Additionally, while it is unclear how well an MPS with interleaved ordering would perform for higher dimensional systems (Gourianov \textit{et al.} consider 3-D simulations using the grouped interwoven geometry, but they refrain from making strong claims about its performance \cite{Gourianov2021quantuminspired}), we expect it to outperform an MPS with sequential ordering. For one, it is no longer possible to have an MPS with S3-like ordering in which dimensions are separated sequentially while also having tensors corresponding to fine grids be grouped together. As such, any higher-dimensional sequentially ordered MPS would likely be susceptible to numerical instability.
Alternatively, as mentioned by Refs.~\cite{Lubasch2018multigrid} and \cite{Gourianov2021quantuminspired}, one might consider representing the data using other tensor network ansatz\"{e}, such as tree tensor networks or 2-D tensor networks (PEPS) \cite{Verstraete2004renormalization, Verstraete2008matrix}. 

In addition to considering higher dimensional systems, there are many other potential directions for future work.
For example, our current implementation uses a basic finite difference scheme to solve the Vlasov equation, and we use explicit RK4 as our time stepping scheme. As a result, despite the exponential speed-up obtained when using the MPS representation to solve the PDE at each time step, the total computational cost still has an exponential scaling due to the CFL time step constraint. To avoid this, we can consider using a semi-Lagrangian method \cite{Cheng1976integration, Kormann2015semilagrangian, Einkemmer2020semilagrangian}. A more involved solution would be to investigate implicit time stepping schemes. Implicit time evolution is often not done because it requires performing a matrix inversion, outweighing the benefits of being able to use a larger time step. However, in the MPS framework, because the solution to the PDE is now represented as a network of smaller tensors, one can consider performing iterative local optimizations to implicitly solve for the next time step \cite{Ripoll2021quantuminspired}. Alternatively, one could consider using the MPS framework with other methods for solving PDEs, such as spectral methods or finite element methods. These methods may also be more robust to noise introduced by the MPS compression methods.


Improving our time evolution scheme and altering the MPS algorithm to conserve plasma properties like energy or momentum, would also be of interest as it may yield more desirable results in some cases. There already exist some algorithms in the tensor train community that we could consider \cite{Einkemmer2018lowrankprojectorsplitting, Einkemmer2021conservation}. Additionally, MPS methods are designed to minimize the L2-norm of the compression error. In contrast, the relevant norm for a distribution function is the L1-norm. Thus, it may also be interesting to investigate how using MPS to represent the square root of the distribution function would compare to the results presented here.

Lastly, as mentioned in the introduction, the MPS methods used here are not specific to the Vlasov-Poisson equation, and extending these methods to solve the Vlasov-Maxwell equations or other kinetic formulations, such as the gyrokinetic equations, is straightforward.

\section{Conclusion}
MPS methods can efficiently represent and solve the 1D1V Vlasov-Poisson equation using a finite-difference scheme. 
We show this by measuring the entanglement entropy and the error generated by compression for numerically exact solutions to the Vlasov-Poisson equation. 
We also perform time evolution with compression at each time step, investigating the behavior in both linear and nonlinear regimes. The success of the MPS method varies depending on the design of the MPS. However, we are ultimately able to compute the dynamics within 10\% accuracy while compressing the distribution function by about a factor of 8. When the solutions are compressed even further, while some details are lost, we are still able to capture general features of the plasma, including the approximate morphology of the electron and ion distribution functions; as well as the oscillation frequencies, linear damping or growth rates, and saturation energies of the electric field. 

\section{Code Availability}
Code is available upon request.

\section{Acknowledgements}
The authors were supported by award DE-SC0020264 from the Department of Energy. The authors also thank Noah Mandell for assisting us in getting started with Gkeyll, and the Gkeyll team for their thorough documentation.

\appendix
\section{Methods}
\subsection{Numerical Experiments}
We demonstrate the utility of the MPS algorithm for solving the Vlasov-Poisson equation through a few test cases. 
All our simulations are done in units normalized to the Debye length $\lambda_D$, electron plasma frequency $\omega_{p,e}$, and the electron thermal velocity $v_{th,e}$.

\subsubsection{Nonlinear Landau damping}
The initial distributions of the ions and electrons are given by
\begin{align}
    f_i(x,v) &= (2\pi v_{th,i}^2)^{-\frac{1}{2}} \, e^{-v^2/2 v_{th,i}^2}, \\ 
    f_e(x,v) &= (2\pi v_{th,e}^2)^{-\frac{1}{2}} \, e^{-v^2/2 v_{th,e}^2} \left(1 + A\cos(kx) \right) ,
\end{align}
where $v_{th,s}^2 = T_s/m_s$ is the thermal velocity of particle species $s$, with $T_s$ the temperature and $m_s$ the mass. In our simulations, we use a realistic mass ratio of $m_i/m_e=1836$, perturbation strength of $A=0.5$, and wavevector $k=0.75$. Simulations are performed on a uniformly discretized grid with periodic boundary conditions in $x$ and zero-gradient boundary conditions in $v$. The bounds of the spatial domain are $\pm a\pi/k$, where $a$ is the number of periods to be included to ensure the grid spacing is as desired. For the electron distribution, the bounds of the simulation domain in $v$ are $\pm 6 v_{th,s}$ for the electron and ion distributions.

\subsubsection{Buneman instability}
The initial distributions of the ions and electrons are given by
\begin{align}
    f_i(x,v) &= (2\pi v_{th,i}^2)^{-\frac{1}{2}} \, e^{-v^2/2 v_{th,i}^2}, \\ 
    f_e(x,v) &= (2\pi v_{th,e}^2)^{-\frac{1}{2}} \, e^{-(v-v_0)^2/2 v_{th,e}^2} \left(1 + A\cos(kx) \right) ,
\end{align}
with $v_0 = \omega_{p,e}/k$.
We use a mass ratio of $m_i/m_e=25$, and perform simulations with perturbation strength of $A=10^{-3}$ and wavevector $k=0.10$. We use a uniform discretization in $x$ from $-\pi/k$ to $\pi/k$ with periodic boundary conditions, and a uniform discretization in $v$ with zero-gradient boundary conditions. For the electron distribution, the bounds of the simulation domain in $v$ are $\pm 6 \omega_{p,e}/k$. For the ion distribution, the bounds are $\pm 50 v_{th,i}$.

\newpage
\subsubsection{Shock-wave formation}
The initial distributions of the ions and electrons are given by
\begin{align}
    f_i(x,v) &= H(x) \exp \left(-(v-v_{0})^2/2 v_{th,i}^2 \right) + \nonumber \\
    & \hspace{0.5cm} (1-H(x)) \exp \left(-(v+v_{0})^2/2 v_{th,i}^2 \right), \\ 
    f_e(x,v) &= H(x) \exp \left(-\left(v-v_{0} \right)^2/2 v_{th,e}^2 \right) + \nonumber \\
    & \hspace{0.5cm} (1-H(x)) \exp \left(-\left(v-v_{0} \right)^2/2 v_{th,e}^2
    \right),
\end{align}
where $H(x)$ is the Heaviside step function, set $v_{0} = 1.5 (T_e/m_i)$, and use a realistic mass ratio of $m_i/m_e=1836$. Our simulations are performed using a uniform discretization in $x$ and $v$, both with zero-gradient boundary conditions. The spatial simulation domain has bounds of $\pm 250 \lambda_D$. For the electron distribution, the bounds of the simulation domain in $v$ are $\pm 6 v_{th,e}$. For the ion distribution, the simulation domain is from $(-10 v_{th,i} - v_{0})$ to $(10 v_{th,i} + v_{0})$.

\subsubsection{Time Evolution Procedure}
To solve the Vlasov-Poisson equations, we compute the spatial advection term using an upwind finite-difference scheme \cite{jardin2010computational} and the gradient in velocity using a centered finite-difference scheme, both with second-order accuracy. Time evolution is performed using standard RK4. We use an adaptive time-stepping scheme, in which the time step is a specified fraction of the maximum allowed by the Courant-Friedrichs-Lewy (CFL) limit \cite{cfl1967partial}. We use a time step that is 0.9 of the CFL limit. The maximum time step is calculated as
\begin{align}
    \Delta t_{\max} =
    \bigg( \sum_s \;\frac{|v_s|_{\max}}{\Delta x} & + \frac{|q_s| |E|_{\max}}{m_s \Delta v_{s}} \nonumber \\
    & \hspace{-0.5cm} + \frac{ \nu_s |v_s|_{\max}}{\Delta v_{s}} + \frac{2 \nu_s v_{th,s}^2}{\Delta v_{s}^2}
    \bigg)^{-1}
    \label{eq:cfl}
\end{align}
where $\Delta x$ and $\Delta v_s$ are the grid discretizations along $x$ and $v$, $|v_s|_{\max}$ is the maximum velocity magnitude, $|E|_{\max}$ is the maximum electric field magnitude (measured at each time step), $v_{th,s}$ is the thermal velocity, and $\nu_s$ is the collision rate.

\subsubsection{Collision Operator}
When collisions are included in the simulations, we use the Dougherty collision operator \cite{dougherty1964model},
\begin{align}
    \mathcal{C}[f_s] = \nu_s \nabla_{\textbf{v},s} \cdot \left[ (\textbf{v}-\textbf{v}_{0,s}) f_s + v_{th,s}^2 \nabla_{\textbf{v},s} f_s \right],
\end{align}
where $\textbf{v}_{0,s}$ is the average velocity, $v_{th,s}$ is the thermal velocity, and $\nu_s$ is the collision frequency.

\begin{widetext}
\section*{Supplemental Information}
\section*{Common Matrix Product State Operations}
In this section we briefly describe the common operations performed on the matrix product states (MPS) when solving the Vlasov-Poisson equation. For a more complete introduction for MPS methods in the context of solving partial differential equations, we direct readers to Refs.~\cite{Lubasch2018multigrid} and \cite{Ripoll2021quantuminspired}. Ref.~\cite{Schollwock2011dmrg} also provides a good overview of MPS methods, even though it is written for applications in quantum physics.

\subsection*{Matrix-vector multiplication}
Assume that the vector is represented as an MPS
\begin{align}
    f(i_1,...i_L) = \sum_{\alpha_1, ..., \alpha_{L-1}} A^{(1)}_{\alpha_1}(i_1) \, A^{(2)}_{\alpha_1,\alpha_2}(i_2) \, ... \, A^{(L)}_{\alpha_{L-1}} (i_L) \, ,
\end{align}
and the operator is represented as an MPO
\begin{align}
    O(o_1,...o_L, i_1,...i_L) = \sum_{\beta_1, ..., \beta_{L-1}} B^{(1)}_{\beta_1}(o_1, i_1) \, B^{(2)}_{\beta_1,\beta_2}(o_2, i_2) \, ... \, B^{(L)}_{\beta_{L-1}} (o_L, i_L) \, .
\end{align}
To perform a matrix-vector multiplication, we contract over the indices $\{i_1,...,i_L\}$, yielding
\begin{align}
    g(o_1,...o_L) = \sum_{\gamma_1, ..., \gamma_{L-1}} C^{(1)}_{\gamma_1}(o_1) \, C^{(2)}_{\gamma_1,\gamma_2}(o_2) \, ... \, C^{(L)}_{\gamma_{L-1}} (o_L) \, ,
\end{align}
where 
\begin{align}
    C^{(n)}_{\gamma_{n-1},\gamma}(o_n) = \sum_{i_n} B^{(n)}_{\beta_{n-1}, \beta_{n}}(o_n,i_n) \, A^{(n)}_{\alpha_{n-1},\alpha} (i_n)
\end{align}
and $\gamma_n = (\alpha_n, \beta_n)$.
Thus, the bond dimension of $g=Af$ is the product of the bond dimensions of $f$ and $A$.

\subsection*{Addition}
Suppose we have two MPS,
\begin{align}
    f(i_1,...i_L) &= \sum_{\alpha_1, ..., \alpha_{L-1}} A^{(1)}_{\alpha_1}(i_1) \, A^{(2)}_{\alpha_1,\alpha_2}(i_2) \, ... \, A^{(L)}_{\alpha_{L-1}} (i_L) \, , \\
    g(i_1,...i_L) &= \sum_{\beta_1, ..., \beta_{L-1}} B^{(1)}_{\beta_1}(i_1) \, B^{(2)}_{\beta_1,\beta_2}(i_2) \, ... \, B^{(L)}_{\beta_{L-1}} (i_L)\, .
\end{align}
The sum of $h=f+g$ is given by 
\begin{align}
    h(i_1,...i_L) &= \sum_{\delta_1, ..., \delta_{L-1}} C^{(1)}_{\delta_1}(i_1) \, C^{(2)}_{\delta_1,\delta_2}(i_2) \, ... \, C^{(L)}_{\delta_{L-1}} (i_L)\, ,
\end{align}
where
\begin{align}
    C^{(n)}(i_n) = 
        \begin{bmatrix} A^{(n)}(i_n) & 0 \\ 0 & B^{(n)}(i_n) \end{bmatrix} \, , 
\end{align}
and $\delta_n$ is the concatenation of indices $\alpha_n$ and $\beta_n$. As such, the bond dimension of $h$ is the sum of the bond dimensions of $f$ and $g$.

\subsection*{Derivatives}
We choose to solve the PDE using finite difference methods on a uniform grid with spacing $\Delta x$. We also choose to map our data onto the MPS with physical dimension $d=2$ using a binary mapping, as described in the main text. 
\subsubsection*{First Derivative, Centered, Periodic Boundary Conditions}
In this case, the first derivative along one axis, assuming periodic boundary conditions and using a second-order centered finite difference scheme, is
\begin{align}
    \frac{\partial}{\partial x} \approx \frac{1}{\Delta x} 
    \begin{bmatrix}
        \mathbb{I} & S^+ + S^- & S^+ + S^- 
    \end{bmatrix}
    \begin{bmatrix}
        \mathbb{I} & S^- & S^+ \\
        0 & S^+ & 0 \\
        0 & 0 & S^-
    \end{bmatrix} 
    \begin{bmatrix}
        \mathbb{I} & S^- & S^+ \\
        0 & S^+ & 0 \\
        0 & 0 & S^-
    \end{bmatrix} 
    \hdots 
    \begin{bmatrix}
        \mathbb{I} & S^- & S^+ \\ 
        0 & S^+ & 0 \\ 
        0 & 0 & S^- \\ 
    \end{bmatrix}
    \begin{bmatrix}
        c_1(S^- - S^+) \\ c_1 S^+ \\ -c_1 S^- 
    \end{bmatrix}
\end{align}
where
\begin{align*}
    S^+ = \begin{bmatrix} 0 & 0 \\ 1 & 0  \end{bmatrix}, \hspace{0.5cm}
    S^- = \begin{bmatrix} 0 & 1 \\ 0 & 0  \end{bmatrix}
\end{align*}
and $c_1 = 1/2$.

\subsubsection*{First Derivative, Forward, Periodic Boundary Conditions}
The second-order accurate forward finite difference MPO is given by
\begin{align}
    \frac{\partial}{\partial x} \approx \frac{1}{\Delta x} 
    \begin{bmatrix}
        \mathbb{I} & S^- & S^+
    \end{bmatrix}
    \begin{bmatrix}
        \mathbb{I} & S^- & 0 \\
        0 & S^+ & 0 \\
        0 & 0 & S^+
    \end{bmatrix} 
    \hdots 
    \begin{bmatrix}
        \mathbb{I} & S^- & 0 \\
        0 & S^+ & 0 \\
        0 & 0 & S^+
    \end{bmatrix} 
    \begin{bmatrix}
        \mathbb{I} & S^- & 0 \\
        0 & S^+ & 0 \\
        0 & S^+ & \mathbb{I} 
    \end{bmatrix}
    \begin{bmatrix}
        c_0 \mathbb{I} + c_1 S^- \\ c_1 S^+ + c_2 \mathbb{I} + c_3 S^- \\ c_3 S^+
    \end{bmatrix}
\end{align}
where $c_0=-3/2, \, c_1=2, \, c_2=-1/2$. 

\subsubsection*{First Derivative, Backward, Periodic Boundary Conditions}
The second-order accurate backward finite difference MPO can be obtained by interchanging $S^+$ and $S^-$, and using $c_0=3/2, \, c_1=-2, \, c_2=1/2$.

\subsubsection*{First Derivative, Centered, Zero Gradient Boundary Conditions}
For central finite difference first derivatives with non-periodic boundary conditions, we compute the first derivative by first building this backbone MPO
\begin{align}
    \frac{\partial}{\partial x} \approx \frac{1}{\Delta x} 
    \begin{bmatrix}
        \mathbb{I} & S^- & S^+
    \end{bmatrix}
    \begin{bmatrix}
        \mathbb{I} & S^- & S^+ \\
        0 & S^+ & 0 \\
        0 & 0 & S^-
    \end{bmatrix} 
    \hdots 
    \begin{bmatrix}
        \mathbb{I} & S^- & S^+ \\ 
        0 & S^+ & 0 \\ 
        0 & 0 & S^- \\ 
    \end{bmatrix}
    \begin{bmatrix}
        c_1(S^- - S^+) \\ c_1 S^+ \\ -c_1 S^- 
    \end{bmatrix}
\end{align}
where $c_1=1/2$. We then add MPOs representing the desired boundary conditions. In the case of zero gradient boundary conditions, we utilize the ghost cell method and prescribe the next grid point outside the simulation boundary to have the same value as the grid point immediately inside of the boundary. Thus, we add the matrices 
\begin{align}
    M_\text{left} & = 
    \begin{bmatrix}  \begin{bmatrix}  
        0 & -c_1  \\ 
        0 & 0
    \end{bmatrix} \end{bmatrix} \otimes 
    \begin{bmatrix} \begin{bmatrix} 0 & 0 \\ 0 & 1  \end{bmatrix} \end{bmatrix}
    \otimes \hdots \otimes
    \begin{bmatrix} \begin{bmatrix} 0 & 0 \\ 0 & 1  \end{bmatrix} \end{bmatrix}
    \, ,
    \\
    M_\text{right} & =
    \begin{bmatrix} \begin{bmatrix} 0 & 0 \\ 0 & 1  \end{bmatrix} \end{bmatrix}
    \otimes \hdots \otimes 
    \begin{bmatrix} \begin{bmatrix} 0 & 0 \\ 0 & 1  \end{bmatrix} \end{bmatrix}
    \otimes
    \begin{bmatrix} \begin{bmatrix} 
        0 & 0 \\
        c_1 & 0 
    \end{bmatrix} \end{bmatrix} 
\end{align}
where $c_1=1/2$ and the MPOs have a total length of $L$.

\subsubsection*{Second Derivative, Centered, Periodic Boundary Conditions}
The MPO for the second-order central finite difference second derivative assuming periodic boundary conditions can be written as
\begin{align}
    \frac{\partial^2}{\partial x^2} \approx \frac{1}{\Delta x^2} 
    \begin{bmatrix}
        \mathbb{I} & S^+ + S^- & S^+ + S^- 
    \end{bmatrix}
    \begin{bmatrix}
        \mathbb{I} & S^+ & S^- \\
        0 & S^- & 0 \\
        0 & 0 & S^+
    \end{bmatrix} 
    \hdots 
    \begin{bmatrix}
        \mathbb{I} & S^+ & S^- \\ 
        0 & S^- & 0 \\ 
        0 & 0 & S^+ \\ 
    \end{bmatrix}
    \begin{bmatrix}
        c_0 \mathbb{I} + c_1(S^- + S^+) \\ c_1 S^- \\ c_1 S^+ 
    \end{bmatrix}
    \, .
\end{align}
where $c_0=-2, c_1=1$.


\subsection*{Dot Product}
In order to take the dot product between two MPS, we have to `diagonalize' one of them into an MPO.
An arbitrary MPS can be written like
\begin{align}
    f(i_1,...i_L) = \sum_{\alpha_1, ..., \alpha_{L-1}} M^{(1)}_{\alpha_1}(i_1) \, M^{(2)}_{\alpha_1,\alpha_2}(i_2) \, ... \, M^{(L)}_{\alpha_{L-1}} (i_L) \, .
\end{align}
Diagonalizing it yields the MPO
\begin{align}
    f(o_1,...,o_L, i_1,...i_L) = \sum_{\alpha_1, ..., \alpha_{L-1}} M^{(1)}_{\alpha_1}(o_1,i_1) \, M^{(2)}_{\alpha_1,\alpha_2}(o_2,i_2) \, ... \, M^{(L)}_{\alpha_{L-1}}(o_L.i_L)
\end{align}
where
\begin{align}
    M_{l,r}(o,i) = M_{l,r}(i) \, \delta_{i, o}
\end{align}
and $\delta_{ij}$ is the Kronecker delta.
Diagrammatically, this would look like
\begin{center}
    {\small

\begin{tikzpicture}[baseline,scale=0.65]
    \begin{scope}[shift={(0.0,0.0)}]
        %
        \draw[thick] (-3,0) -- (2,0);
        \draw[thick] (-3,0) -- (-3,0.75);
        \draw[thick] (-2,0) -- (-2,0.75);
        \draw[thick] (-1,0) -- (-1,0.75);
        \draw[thick] ( 0,0) -- ( 0,0.75);
        \draw[thick] ( 1,0) -- ( 1,0.75);
        \draw[thick] ( 2,0) -- ( 2,0.75);
        \draw[thick, blue, rounded corners] (-3,0.5) -- (-3.4,0.5) -- (-3.4, -0.5) -- (-3.0,-0.5) -- (-3.0,-1.0);
        \draw[thick, blue, rounded corners] (-2,0.5) -- (-2.4,0.5) -- (-2.4,0.1);
        \draw[thick, blue, rounded corners] (-2.4,-0.1) -- (-2.4, -0.5) -- (-2.0,-0.5) -- (-2.0,-1.0);
        \draw[thick, blue] (-2.4,0.1) arc (90:270:0.1);
        \draw[thick, blue, rounded corners] (-1,0.5) -- (-1.4,0.5) -- (-1.4,0.1);
        \draw[thick, blue, rounded corners] (-1.4,-0.1) -- (-1.4, -0.5) -- (-1.0,-0.5) -- (-1.0,-1.0);
        \draw[thick, blue] (-1.4,0.1) arc (90:270:0.1);
        \draw[thick, blue, rounded corners] (0,0.5) -- (-0.4,0.5) -- (-0.4,0.1);
        \draw[thick, blue, rounded corners] (-0.4,-0.1) -- (-0.4, -0.5) -- (0,-0.5) -- (0,-1.0);
        \draw[thick, blue] (-0.4,0.1) arc (90:270:0.1);
        \draw[thick, blue, rounded corners] (1,0.5) -- (0.6,0.5) -- (0.6,0.1);
        \draw[thick, blue, rounded corners] (0.6,-0.1) -- (0.6, -0.5) -- (1.0,-0.5) -- (1.0,-1.0);
        \draw[thick, blue] (0.6,0.1) arc (90:270:0.1);
        \draw[thick, blue, rounded corners] (2,0.5) -- (1.6,0.5) -- (1.6,0.1);
        \draw[thick, blue, rounded corners] (1.6,-0.1) -- (1.6, -0.5) -- (2.0,-0.5) -- (2.0,-1.0);
        \draw[thick, blue] (1.6,0.1) arc (90:270:0.1);
        \draw[thick,fill=black!30!green] (-3.0,0) circle (0.25);
        \draw[thick,fill=black!30!green] (-2.0,0) circle (0.25);
        \draw[thick,fill=black!30!green] (-1.0,0) circle (0.25);
        \draw[thick,fill=black!30!green] ( 0.0,0) circle (0.25);
        \draw[thick,fill=black!30!green] ( 1.0,0) circle (0.25);
        \draw[thick,fill=black!30!green] ( 2.0,0) circle (0.25);%
    \end{scope}
\end{tikzpicture}

}    
\end{center}
where the squiggly lines in blue represents the delta function. (The blip as it crosses the horizontal bonds means that it does not interact with it.)  In this new tensor network, each tensor now has two legs, and thus it has the form of a matrix product operator. 

\subsection*{Inverse Laplacian}
Having an MPO for the second derivative and a method with which to add MPOs together, we can obtain an MPO representation of the Laplacian. To obtain the inverse, one can use Newton's method, which solely involves matrix multiplications. Alternatively, we can use a density matrix renormalization group style optimization. 

However, because we only consider 1D1V systems in this work, we obtain the operator by starting with a matrix representation of the Laplacian operator, inverting it, and then converting it into an MPO. For higher dimensional systems, this method still may be a viable option because even though the computation is expensive, the inverse Laplacian can be stored after it is computed and thus only needs to be computed once.

For 1-D systems, we found that the MPO of the inverse Laplacian (where the Laplacian is accurate to second order) requires a bond dimension of about 5 if retaining singular values up to $10^{-10}$.

\subsection*{Upwind time evolution}
For the advection term in the Vlasov equation, we use an upwind time evolution scheme, in which gradients multiplied by positive velocities are computed using a backward finite difference stencil while gradients multiplied by negative velocities are computed using a forward finite difference stencil, or 
\begin{align}
   \frac{\partial}{\partial x} \approx \left[ \frac{\partial}{\partial x} \right]_{\text{forward}} H(v) + \left[ \frac{\partial}{\partial x} \right]_\text{backward} (1-H(v))
\end{align}
where $H(v)$ is the Heaviside step function

MPOs of the derivatives using a forward and backward finite difference stencil are described above. The MPO representing the Heaviside function can be obtained by numerically solving for the MPS representation of $H(v)$, and then diagonalized into an MPO. For a discretized grid centered at 0, the bond dimension of the MPO is 2. If the MPOs of the forward and backward time evolution derivative have bond dimension $D_f$ and $D_b$ respectively, then the new MPO will have at most a bond dimension $D_f$+$D_b$. Thus, while there is some overhead associated with performing an upwind scheme, if it results in less noise in the time evolved state, it is ultimately worthwhile, especially since the noise can often artificially reduce the compressibility of the MPS.

\subsection*{Integration}
A first-order integration scheme (required for solving the Poisson equation),
\begin{align}
    \int_{x_0}^{x_{N-1}} f(x) \, dx \approx \sum_{i=0}^{N-1} f(x_i) \Delta x
\end{align}
can be performed by contracting the MPS representing $f(x)$ with the MPS
\begin{align}
    I(i_1,...,i_L) = \Delta x
    \begin{bmatrix} 1 \\ 1 \end{bmatrix} \otimes 
    \begin{bmatrix} 1 \\ 1 \end{bmatrix} \otimes \hdots \otimes
    \begin{bmatrix} 1 \\ 1 \end{bmatrix}
\end{align}

\section*{Canonicalization and Compression}
An important part of MPS algorithms is the compression of the MPS to a smaller bond dimension. Doing the compression accurately requires one to put the MPS in a canonical form.


An MPS has a left canonical form and a right canonical form, respectively defined as
\begin{align}
    f(i_1,...,i_L) &= \sum_{\alpha_1,...,\alpha_{L-1}} L_{\alpha_1}^{(1)} (i_1) \,  ... \, L_{\alpha_{L-2},\alpha_{L-1}}^{(L-1)} (i_{L-1}) \, M_{\alpha_{L-1}}^{(L)} (i_L) \, , \\
    f(i_1,...,i_L) &= \sum_{\alpha_1,...,\alpha_{L-1}} M_{\alpha_1}^{(1)} (i_1) \, R_{\alpha_1,\alpha_2}^{(2)}(i_2) \, ... \, R_{\alpha_{L-1}}^{(L)} (i_L) \, ,
\end{align}
and a mixed canonical form where the tensors left of the $n^\text{th}$ tensor are in left canonical form and tensors to the right are in right canonical form:
\begin{align}
    f(i_1,...,i_L) = \sum_{\alpha_1,...,\alpha_{L-1}} L_{\alpha_1}^{(1)} (i_1) \, ... \, L_{\alpha_{n-2},\alpha_{n-1}}^{(n-1)} (i_{n-1}) \, M_{\alpha_{n-1},\alpha_{n}}^{(n)}(i_n) \, R_{\alpha_n,\alpha_{n+1}}^{(n+1)}(i_{n+1}) \, ... \,
    R_{\alpha_{L-1}}^{(L)}(i_{L})
\end{align}
where the tensors $L_{l,r}(u)$ and $R_{l,r}(u)$ have the property
\begin{align}
    \sum_{l,u} L_{l,r} (u) L^*_{l,r'} (u) = \delta_{r,r'} \quad, \quad
    \sum_{r,u} R_{l,r} (u) R^*_{l',r} (u) = \delta_{l,l'} 
\end{align}
while the tensors $M_{l,r}(u)$ have no constraints.

To canonicalize the MPS, one performs an iterative QR decomposition procedure. For example, to put the MPS in left canonical form, starting from the left-most tensor in the chain, we decompose the $n^\text{th}$ tensor into two using QR decomposition, and then contract the $R$ matrix into the $(n+1)^\text{th}$ tensor:
\begin{center}
    {\small

\begin{tikzpicture}[baseline,scale=0.65]
    \node at (-2,2) {QR canonicalization};
    \begin{scope}[shift={(-1,0)}]
        \node at (-2.25,0) {$\hdots$};
        \draw[thick] (-1.75,0) -- ( 0.75,0);
        \node at ( 1.25,0) {$\hdots$};
        \draw[thick] (-1,0) -- (-1,0.75);
        \draw[thick] ( 0,0) -- ( 0,0.75);
        %
        \draw[thick,fill=black!20!yellow] (-1.0,0) circle (0.25);
        \draw[thick,fill=black!30!green] ( 0.0,0) circle (0.25);
        \node at (2.5,0) {\large $\Rightarrow$};
    \end{scope}
    \begin{scope}[shift={(5,0)}]
        \node at (-2.25,0) {$\hdots$};
        \draw[thick] (-1.75,0) -- ( 1.75,0);
        \node at ( 2.25,0) {$\hdots$};
        \draw[thick] (-1,0) -- (-1,0.75);
        \draw[thick] ( 1,0) -- ( 1,0.75);
        \draw[fill=black!20!yellow] 
                (-1.15, 0.3)--(-0.65,-0.0)--(-1.15,-0.3)--cycle;
        \draw[thick,fill=black!20!yellow] ( 0.0,0) circle (0.25);
        \draw[thick,fill=black!30!green] ( 1.0,0) circle (0.25);
        \node at (-1.0,-1.0) {$Q$};
        \node at ( 0.0,-1.0) {$R$};
        \draw[dashed] (0.5,0) ellipse (1.0 and 0.5);
        \node at (3.0,0) {\large $\Rightarrow$};
    \end{scope}
    \begin{scope}[shift={(11,0)}]
        \node at (-2,0) {$\hdots$};
        \draw[thick] (-1.5,0) -- ( 0.5,0);
        \node at ( 1,0) {$\hdots$};
        \draw[thick] (-1,0) -- (-1,0.75);
        \draw[thick] ( 0,0) -- ( 0,0.75);
        %
        \draw[fill=black!20!yellow] 
                (-1.15, 0.3)--(-0.65,-0.0)--(-1.15,-0.3)--cycle;
        \draw[thick,fill=black!30!green] ( 0.0,0) circle (0.25);
    \end{scope}
\end{tikzpicture}

}
\end{center}
\begin{align}
    \sum_{\alpha_1,\alpha_2,\hdots} M_{\alpha_1}^{(1)}(i_1) \, M_{\alpha_1,\alpha_2}^{(2)}(i_2) \hdots 
    & = \sum_{\alpha_1, \alpha_2, \hdots} \left( \sum_{\beta_1} Q_{\beta_1}^{(1)}(i_1) \, R_{\beta_1,\alpha_1}^{(1)} \right) M_{\alpha_1,\alpha_2}^{(2)}(i_2) \hdots \nonumber \\
    & = \sum_{\beta_1,\alpha_2,\hdots} Q_{\beta_1}^{(1)}(i_1) \left( \sum_{\alpha_1} R_{\beta_1,\alpha_1}^{(1)} \,  M_{\alpha_1,\alpha_2}^{(2)}(i_2) \right) \hdots 
    = \sum_{\beta_1,\alpha_2,\hdots} Q_{\beta_1}^{(1)}(i_1) \, \tilde{M}_{\beta_1, \alpha_2}^{(2)}(i_2)\hdots 
\end{align}
The orthonormal properties of $Q$ ensures that it is of left canonical form, which we represent using a left-facing triangle. One repeats this process with the following tensor until one reaches the end of the MPS.

Once the MPS is in canonical form, we can compress the MPS using a similar iterative scheme, except instead of performing a QR decomposition one decomposes the tensor via singular value decomposition (SVD) and retains only the $D$ largest singular values. In the cartoon below, we assume the MPS is already in right canonical form. We decompose the $n^\text{th}$ tensor using SVD. We reduce the size of the bonds between $U$, $S$ and $V^T$ by keeping only the $D$ largest singular values and the corresponding orthonormal vectors. Again, the $U$ tensor by definition is in left canonical form, so we contract $S$ and $V^T$ into the $(n+1)^\text{th}$ tensor in the MPS.
\begin{center}
    {\small 

\begin{tikzpicture}[baseline,scale=0.65]
    \node at (-2,2) {SVD compression};
    \begin{scope}[shift={(-1,0)}]
        \node at (-2.25,0) {$\hdots$};
        \draw[thick] (-1.75,0) -- ( 0.75,0);
        \node at ( 1.25,0) {$\hdots$};
        \draw[thick] (-1,0) -- (-1,0.75);
        \draw[thick] ( 0,0) -- ( 0,0.75);
        %
        \draw[thick,fill=black!20!yellow] (-1.0,0) circle (0.25);
        \draw[fill=black!30!green] 
                ( 0.15, 0.3)--(-0.35,-0.0)--( 0.15,-0.3)--cycle;
        \node at (2.5,0) {\large $\Rightarrow$};
    \end{scope}
    \begin{scope}[shift={(5,0)}]
        \node at (-2.25,0) {$\hdots$};
        \draw[thick] (-1.75,0) -- ( 2.75,0);
        \node at ( 3.25,0) {$\hdots$};
        \draw[thick] (-1,0) -- (-1,0.75);
        \draw[thick] ( 2,0) -- ( 2,0.75);
        \draw[fill=black!20!yellow] 
                (-1.15, 0.3)--(-0.65,-0.0)--(-1.15,-0.3)--cycle;
        \draw[thick,fill=black!20!yellow] ( 0.0,0) circle (0.25);
        \draw[fill=black!20!yellow] 
                ( 1.15, 0.3)--( 0.65,-0.0)--( 1.15,-0.3)--cycle;
        \draw[fill=black!30!green] 
                ( 2.15, 0.3)--( 1.65,-0.0)--( 2.15,-0.3)--cycle;
        %
        \node at (-1.0,-1.0) {$U$};
        \node at ( 0.0,-1.0) {$S$};
        \node at ( 1.0,-1.0) {$V^T$};
    \end{scope}
    \begin{scope}[shift={(13,0)}] 
        \node at (-3.5,0) {\large $\approx$};
        \node at (-2.25,0) {$\hdots$};
        \draw[thick] (-1.75,0) -- ( 2.75,0);
        \node at ( 3.25,0) {$\hdots$};
        \draw[thick] (-1,0) -- (-1,0.75);
        \draw[thick] ( 2,0) -- ( 2,0.75);
        \draw[fill=black!20!yellow] 
                (-1.15, 0.3)--(-0.65,-0.0)--(-1.15,-0.3)--cycle;
        \draw[thick,fill=black!20!yellow] ( 0.0,0) circle (0.25);
        \draw[fill=black!20!yellow] 
                ( 1.15, 0.3)--( 0.65,-0.0)--( 1.15,-0.3)--cycle;
        \draw[fill=black!30!green] 
                ( 2.15, 0.3)--( 1.65,-0.0)--( 2.15,-0.3)--cycle;
        %
        \node at (-1.2,-1.0) {$\left[U\right]_D$};
        \node at ( 0.0,-1.0) {$\left[S\right]_D$};
        \node at ( 1.2,-1.0) {$\left[V^T\right]_D$};
        \draw[dashed] (1.0,0) ellipse (1.5 and 0.6);
        \node at (4.0,0) {\large $\Rightarrow$};
        \node at (-0.5, 0.35) {\small $D$};
        \node at ( 0.5, 0.35) {\small $D$};
    \end{scope}
    \begin{scope}[shift={(20,0)}] 
        \node at (-2,0) {$\hdots$};
        \draw[thick] (-1.5,0) -- ( 0.5,0);
        \node at ( 1,0) {$\hdots$};
        \draw[thick] (-1,0) -- (-1,0.75);
        \draw[thick] ( 0,0) -- ( 0,0.75);
        %
        \draw[fill=black!20!yellow] 
                (-1.15, 0.3)--(-0.65,-0.0)--(-1.15,-0.3)--cycle;
        \draw[thick,fill=black!30!green] ( 0.0,0) circle (0.25);
        \node at (-0.5, 0.35) {\small $D$};
    \end{scope}
\end{tikzpicture}

}
\end{center}
\begin{align}
    \sum_{\alpha_1,\alpha_2,\hdots} M_{\alpha_1}^{(1)}(i_1) \, R_{\alpha_1,\alpha_2}^{(2)}(i_2) \hdots 
    & = \sum_{\alpha_1, \alpha_2, \hdots} \left( \sum_{\beta_1} U_{\beta_1}^{(1)}(i_1) \, S_{\beta_1} \, (V^T)_{\beta_1,\alpha_1}^{(1)} \right) R_{\alpha_1,\alpha_2}^{(2)}(i_2) \hdots \nonumber \\
    & = \sum_{\beta_1,\alpha_2,\hdots} U_{\beta_1}^{(1)}(i_1) \left( \sum_{\alpha_1} S_{\beta_1} \, (V^T)_{\beta_1,\alpha_1}^{(1)}  R_{\alpha_1,\alpha_2}^{(2)}(i_2) \right) \hdots \\[0.2cm]
    & = \sum_{\beta_1,\alpha_2,\hdots} U_{\beta_1}^{(1)}(i_1) \, \tilde{M}_{\beta_1, \alpha_2}^{(2)}(i_2) \, R_{\alpha_2,\alpha_3}^{(3)}(i_3) \hdots 
\end{align}
Again, we represent the left/right canonical tensors with left/right-facing triangles. This procedure is repeated for the $(n+1)^\text{th}$ tensor until one reaches the right end of the chain.

An alternative compression scheme (not used in this work) is a local optimization scheme inspired by density matrix renormalization group (DMRG) methods. Here, one sweeps through the tensors and updates them such that the error between the original MPS $f$ and the compressed result $f'$ is minimized. For example, the optimal value of the $n^\text{th}$ tensor ($M^{(n)}$) of the compressed MPS is found by solving
\begin{align}
    \arg \min_{M^{(n)}} ||f - f'||^2 \, ,
\end{align}
which can be done using standard methods like conjugate gradient descent.
Because the updates are local, one typically has to sweep through the MPS at least a few times to converge to the optimal solution. The advantage of optimization-based compression is that one now can impose constraints on the system (as is done in Ref.~\cite{Gourianov2021quantuminspired}).



\subsection*{Modified Compression Scheme}
The technical details of compression via SVD are outlined above. We mention that one typically compresses the MPS by performing SVD decompositions starting from one end and ending at the other. However, in an effort to minimize the introduction of unphysical noise as a result of state compression at each time step, we use an alternative truncation scheme inspired by tensor train methods for the sequentially ordered MPS. 

In our 2D system, the middle bond corresponds to entanglement between data along $x$ and $v$. When bipartitioning the MPS at this bond, we can write the distribution function as \[f(x,v; t) = \sum_r \sigma_r(t) g_r(x; t) h_r(v; t) \] where $g_r(x; t)$ and $h_r(v; t)$ are orthonormal 1-D functions. Keeping the $D$ largest singular values is equivalent to retaining only the terms in the sum corresponding to the largest coefficients $\sigma_r$. This low-rank approximation should not generate unphysical noise within the functions $g_r(x;t)$ and $h_r(v;t)$ themselves. In contrast, compressing the other bonds, which carry information between coarser and finer grid resolutions within $x$ and $v$, could result in numerical artifacts that cause numerical instabilities. 

So, in our modified compression scheme we first compress the center bond joining the two halves corresponding to the two different dimensions. After performing this compression step and removing some information from our state, we then continue to compress $g_r(x;t)$ and $h_r(v;t)$ by compressing the remaining tensors in the MPS. The steps of the compression are given below:
\begin{enumerate}
    \item Write the MPS in mixed canonical form with the with tensors left of $L+1$ in left canonical form and the tensor right of $L+1$ in right canonical form.
    \item Decompose the tensor via singular value decomposition $M^{(L+1)} \rightarrow U S V^T$
    \item To avoid extra canonicalization steps, we insert diagonal matrices $S^{-1}$ and $S$ in between $S$ and $V^T$
    \item Contract $L^{(L)} U S \rightarrow M^{(L)}$
    \item Contract $S V^T R^{(L+2)} \rightarrow M^{(L+2)}$
    \item Perform iterative SVD compression scheme for tensors at positions $L$  to $1$
    \item Perform iterative SVD compression scheme for tensors at positions $L+1$ to $2L$
    \item Absorb $S^{-1}$ into a neighboring tensor. 
\end{enumerate}
Note that the compressed MPS is no longer in a canonical form. However, this does not matter since we immediately apply MPOs to the MPS, which also does not preserve canonicalization.

Interestingly, we found that this compression scheme can improve performance for S1 or S2 ordering, but does not seem to significantly affect S3 results.

\subsection*{Compressed time evolution algorithm}
Given an initial value problem $\frac{\partial f}{\partial t} = y(f,t)$, the fourth order Runge-Kutta time evolution scheme is given as follows:
\begin{align}
    f_{n+1} = f_n \, + \, & \frac{1}{6} \Delta t \left(k_1 + 2k_2 + 2k_3 + k_4 \right) \\
    \text{where} \quad k_1 & = y\left(t_n, f_n\right) \\
    k_2 & = y\left(t_n+\frac{\Delta t}{2}, f_n + \frac{\Delta t}{2} k_1\right) \\
    k_3 & = y\left(t_n+\frac{\Delta t}{2}, f_n + \frac{\Delta t}{2} k_2\right) \\
    k_4 & = y\left(t_n+\Delta t, f_n + \Delta t k_3\right)
\end{align}
The most accurate way of computing the next time step would be to perform no active compression until obtaining the state at the next time step, $f_{n+1}$. However, while this algorithm formally still would scale like $\mathcal{O}(D^3 \log(N))$, the constant scaling factor may be large. 

However, we could consider performing intermediate compression steps, such as compressing the intermediate states as well as the derivatives.
We investigate different levels of compression, defined as 
\begin{enumerate}
    \setlength{\itemsep}{5pt}
    \setlength{\parskip}{0pt}
    \setlength{\parsep}{0pt}
    \item[C1.] Only compression of the state at the next time step $f_{n+1}$
    \item[C2.] Additionally compress intermediate states, e.g. $f_n + \frac{\Delta t}{2} k_1$
    \item[C3.] Additionally compress the sum of derivatives ($k_1 + 2k_2 + 2k_3 + k_4$) 
    \item[C4.] Additionally compress each of the derivatives
    \item[C5.] Additionally compress terms that are added together when computing the derivatives
\end{enumerate}
In general, as one goes down the list, we expect the compression scheme to be cheaper but also less accurate. While the later schemes actively compress the MPS more times, they are cheaper because (1) due to the $\mathcal{O}(D^3)$ scaling of computational costs, performing multiple compressions of MPS with smaller bond dimension is often cheaper than performing a single compression of an MPS with larger bond dimension, and (2) even if the MPS is not actively compressed to the specified bond dimension at the intermediate steps, we still perform the MPS compression procedure and only truncate singular values such that the truncation error at each bond is less than $10^{-10}$ and one does not restrict the MPS bond dimension. Otherwise, the bond dimension of our state would quickly become unmanageable.

We find that in the case of shock-wave formation, compression at each intermediate state in the RK4 time stepping scheme performs comparably to only compressing the final state of the next time step. In contrast, compressing the derivatives introduces significantly more error. This is shown in Fig.~\ref{fig:SI_compression}.

\begin{figure}[h]
    \centering
    \includegraphics[width=\linewidth,trim={0cm 0cm 0cm 0}, clip]{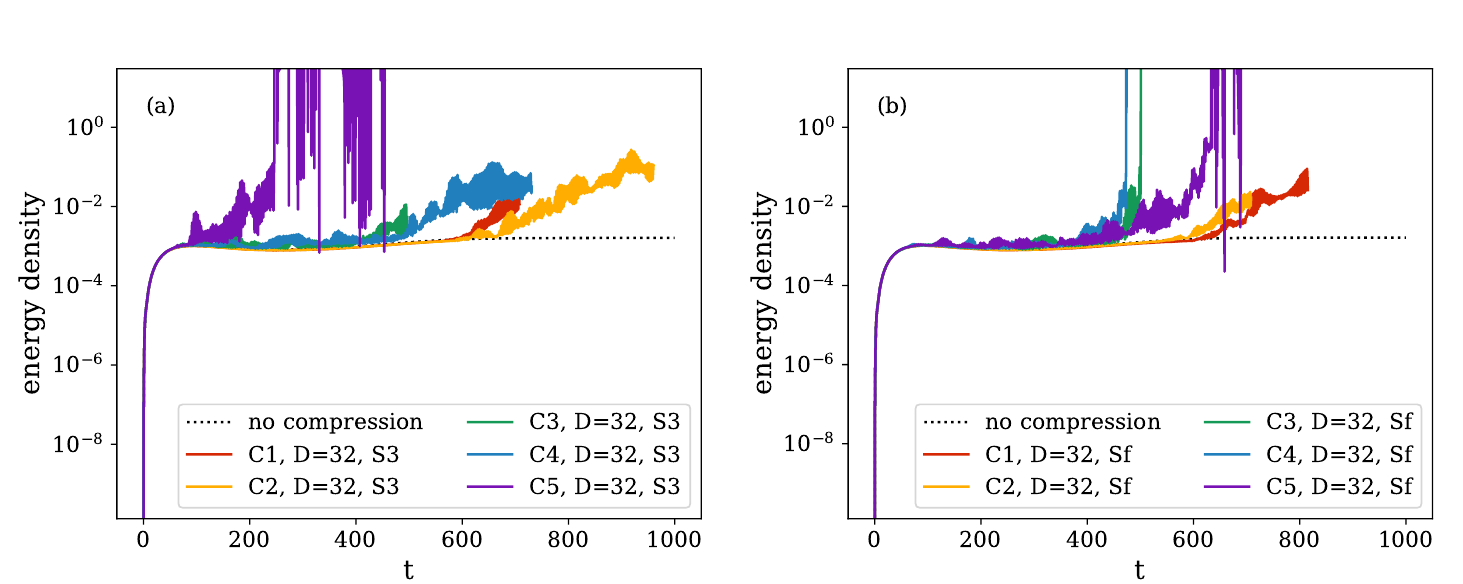}
    \caption{Electric field energy densities the for shock-wave formation test case using the different compressed time evolution schemes detailed above. Results are for (a) S3 and (b) S1 orderings. }
    \label{fig:SI_compression}
\end{figure}

\section*{Additional results for the Buneman instability}
In Fig.~\ref{fig:SI_bun_dists}, we compare the ion and electron distribution functions obtained without compression, with compression to bond dimension $D=64$ at each time step, and with compression to bond dimension $D=32$. We also show results obtained using Gkeyll for reference.

In Fig.~\ref{fig:SI_bun_moments}, we plot the density, momentum, and energy of the ion and electron distribution functions. The uncompressed result and the $D=64$ result show good agreement with each other. They also show reasonable agreement with results from Gkeyll at shorter simulation times, but the differences become more significant at longer times, especially for the measurement of electron momentum. 

\pagebreak

\begin{figure}[h]
    \centering
    \includegraphics[width=\linewidth]{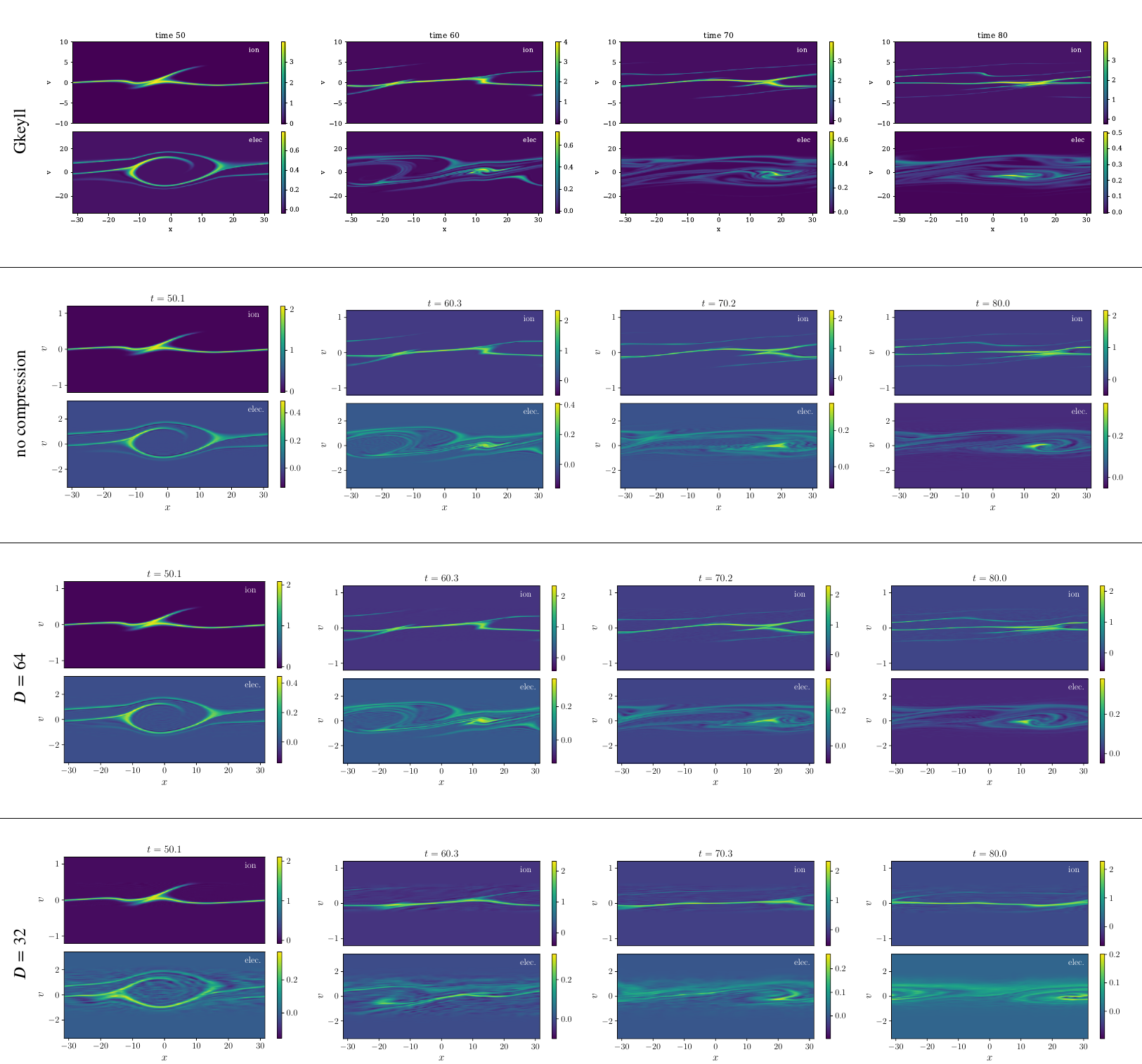}
    \caption{Ion and electron distributions at specified times for the Buneman instability with initial perturbation of wavevector $k=0.10$ and amplitude $A=10^{-3}$. The different rows are results obtained with Gkeyll~\cite{Juno2018discontinuous}, our code with no compression, and our code with compression to $D=64$ and $D=32$ at each time step (using S3 ordering). }
    \label{fig:SI_bun_dists}
\end{figure}

\pagebreak

\begin{figure}[h]
    \centering
    \includegraphics[width=\linewidth]{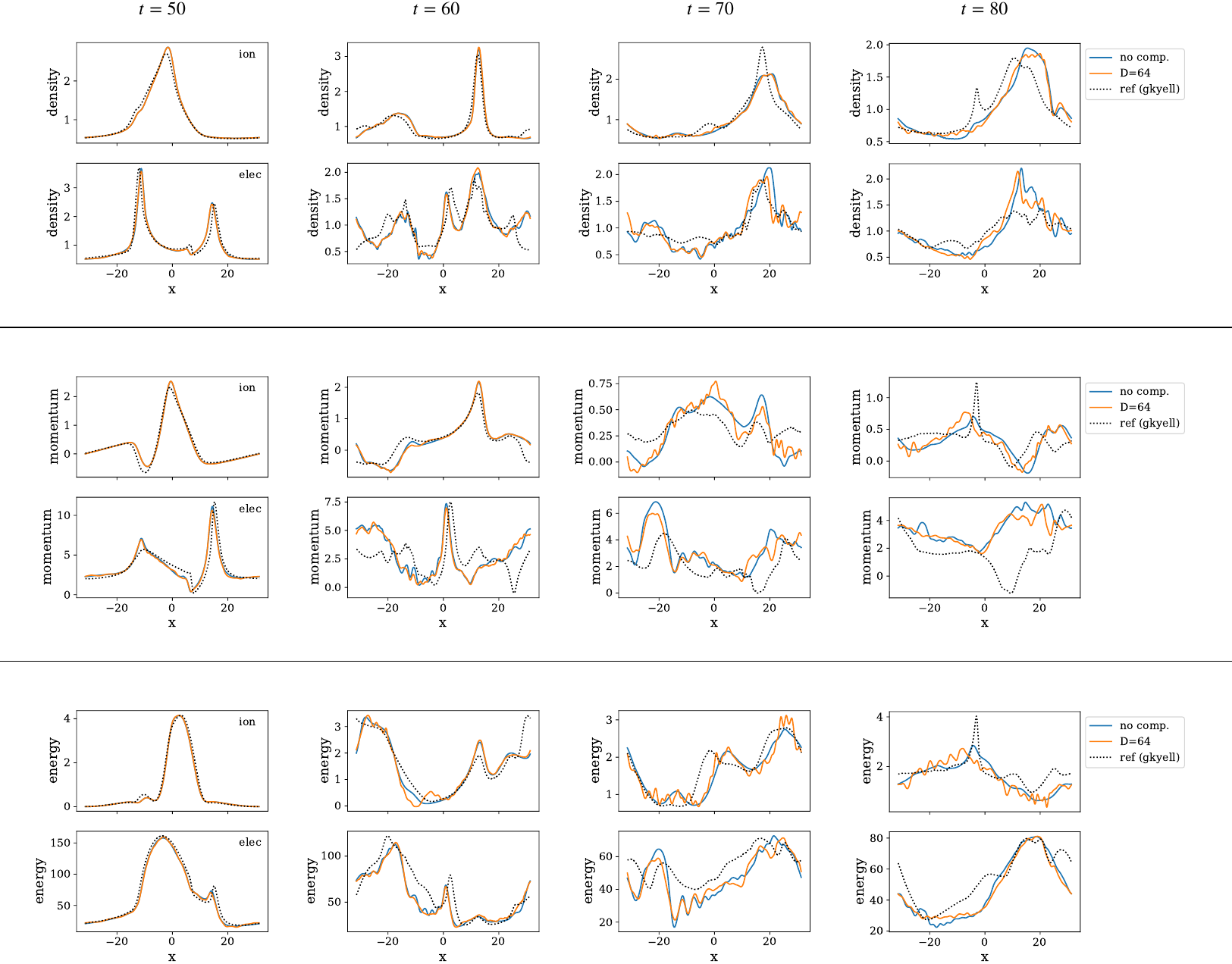}
    \caption{Zeroth, first and second moments of the distribution functions for the Buneman instability with initial perturbation of wavevector $k=0.10$ and amplitude $A=10^{-3}$ at the specified times. Plots compare results obtained without compression, with compression to $D=64$ at each time step, and from Gkeyll. MPS results are obtained using the S3 ordering. Plots on the top and bottom of each row correspond to the ion and electron distributions, respectively.}
    \label{fig:SI_bun_moments}
\end{figure}

\end{widetext}

\pagebreak

\bibliographystyle{apsrev4-1}
\bibliography{pnas-sample}

\end{document}